\documentclass{article}
\usepackage[left=3cm,right=3cm,top=1.5cm,bottom=2cm]{geometry} 
\usepackage{amsmath} 
\usepackage{graphicx}
\usepackage{subcaption}
\usepackage{stackengine}
\usepackage{float}

\begin{document}

\newtheorem{defi}{Definition}
\newtheorem{theo}{Theorem}
\newtheorem{prop}{Proposition}
\newtheorem{lem}{Lemma}
\newtheorem{cor}{Corollary}
\newtheorem{refin}{Refinement}

\title{The bounded 19-vertex model}
\author{Kari Eloranta\\
  University of Helsinki\\
{\tt kari.v.eloranta@gmail.com}}

\date{\today}
\maketitle

\begin{abstract}
\noindent We study the 19-vertex model of Statistical Mechanics in a square with the domain wall boundary condition. Using the minimal set of generating flip actions we build a parametrized dynamic version of the model. For all observed dynamic weight values the equilibrium states exhibit clear limit shapes. Although the model in a way incorporates the 6-vertex model, the reason for the existence of the limit shape is fundamentally more general. We conclude with a further study relating the local path geometry to the macroscopic shape geometry. 
\vskip .05truein  
\noindent Keywords: Vertex model, limit shape, arctic curve
\end{abstract}

\section{Introduction}

Vertex models of Statistical Mechanics are usually defined on a lattice, sometimes on a graph. Instead of spin variables at vertices one specifies arrow orientations on the edges. The {\bf vertex rule} determines which arrow arrangements are allowed at a vertex. Once this is globally satisfied one has a configuration.

Perhaps the best known such model is the 6-vertex/Ice model defined on the square lattice (\cite{B}). Its vertex rule requires that at every vertex there are two incoming and two outgoing arrows. In this study we consider its generalization, the 19-vertex model, where the vertex rule posits {\bf equal number of incoming and outgoing arrows}. The legal vertex configurations with multiplicities (which account for rotations and reflections and sum up to 19) are then 

\setcounter{figure}{0}
\begin{figure}[H]
\centerline{\includegraphics[height=2cm]{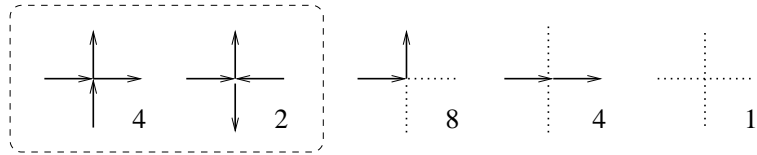}}
\caption{19-vertex rule: allowed vertex configurations with multiplicities.}
\end{figure}

\noindent While the dotted edges indicate the absence of arrow, the frame on the left singles out the vertex configurations allowed in 6-vertex model. Since the 19-vertex rule is less restricting its configuration set will be a strict superset of that of 6-vertex model.

Properties of the 19-vertex model on the infinite square lattice, infinite cylinder and torus, twisted torus etc. has been studied in past (see e.g. \cite{IK},\cite{PW},\cite{H}). Here we will concentrate on behavior of the model on a bounded domain for a very specific reason.

With a suitable boundary condition on a bounded domain many tilings and lattice models exhibit striking {\bf limit shape} behavior in the scaling limit. Generically configurations then show a strict separation of near boundary frozen and macroscopically off-boundary temperate phases and perhaps even further in the interior a gas phase. Apart from obvious visual clues the phases can be characterized with qualitatively different decay rates of correlations. The celebrated first demonstration of this was with dimers on an Aztec diamond (\cite{JPS}). Later the phenomenon has been found and analyzed to a varying degree of rigor e.g. in the context of lozenge and fortress tilings, groves, Young-diagrams, Alternating Sign Matrices and Statistical Mechanics models including the Ice-model on various lattices (see e.g. \cite{KO},\cite{PS},\cite{V},\cite{CP},\cite{E2}).

\vskip .1truein
\noindent The purpose of this paper is first to investigate whether the limit shape phenomenon shows up in 19-vertex model. We will use domain and boundary condition familiar from Ice context (leading to the limit shape phenomena there) but since 19-vertex configurations lack some of the rigidity of Ice configurations, nothing is given from the outset. Our study utilizes the dynamic version of the model, the equilibrium of which corresponds to the usual static model.

To summarize our simulation results the 19-vertex model seems to exhibit limit shape behavior for the domain wall boundary condition in a square for a large range of parameter values (dynamic weights). What is even more interesting is that it does not need an embedded Ice-model to do this. There is a way of factoring out the Ice-dynamics from the dynamic 19-vertex model, yet still in the equilibrium a highly non-trivial limit shape shows up. Indeed the results seem to indicate that the geometry of the limit shape is determined by only subaction(s) of the entire 19-vertex dynamics. We also indicate a parametrization of the local path geometry and how this in turn influences in a rather striking way in the macroscopic limit shape.

\section{Set-up}

We will study the 19-vertex model on a square in ${\rm {\bf Z}}^2$ with fixed boundary configuration. As in the Ice context any off-boundary fully oriented and unidirectional loop can be reversed this action resulting in a new legal configuration for the given boundary condition. This is all one can do to perturb an Ice-configuration and one can show that there this action generates all legal configurations (from a seed configuration). In Ice-context one can in fact restrict to unidirectional 1-loops (arrows around a unit square) and just these generate the full configuration set. This is an example of an {\bf elementary action/move} or more colloquially a {\bf flip}. For dimers a generating flip rotates a dimer pair ($2\times 2$ arrangement) by $\frac{\pi}{2}.$

For 19-vertex model one needs several different types of flips (together with all their rotations and reflections) to generate. The minimal set is shown in Figure 2. Type III has three subactions within it; subsequently we will distinguish between them depending on the part of the parameter space we are in. The subaction in the frame on the right is the only one needed to generate all Ice configurations. 

\setcounter{figure}{1}
\begin{figure}[H]
\centerline{\includegraphics[height=5cm]{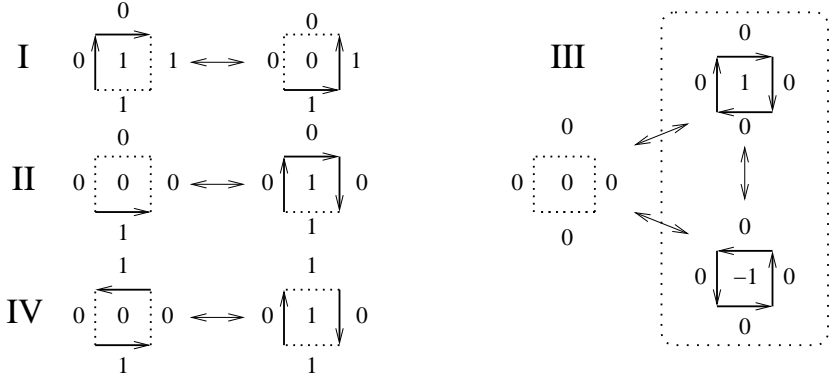}}
\caption{Minimal set of elementary actions and heights around them.}
\end{figure}

\noindent {\bf Height} is a function from to the dual lattice ${\rm {\bf Z}}^2+\left(\frac{1}{2},\frac{1}{2}\right)$ to the integers. Moving from neighboring dual lattice point to another the height increases by one if we cross a left-pointing configuration arrow, decreases by one if the crossed arrow is right-pointing and remains constant if no arrow is crossed. Given a configuration the height function defines a discrete surface over it which is unique upto an additive constant.

The Arabic numbers in the Figure indicate height around and inside the \lq\lq elementary configurations\rq\rq. One readily deduces that these are indeed all the local configurations that can ever be acted upon in a 19-vertex configuration.

To have a dynamic model we need the irreducibility of the configuration set under this minimal collection of elementary moves:

\begin{theo}
Given a 19-vertex configuration on a bounded domain, any other legal configuration with the same boundary condition can be generated from the former using a finite sequence of elementary actions I-IV. A strict subset of actions will not suffice.
\end{theo}

\noindent The proof utilizing a lexicographic sweep over the domain (as for Ice in \cite{E1}) does not readily seem to generalize here but one can argue alternatively using the height. With a sequence of elementary moves acting on local height minima one can build a finite chain of ascending configurations from any configuration to the highest one corresponding to the given boundary (using height one can introduce a natural partial order on the configurations). Hence two such flip chains joined at the maximal element connects any two configurations. That none of the moves I-V can be omitted can be argued with counterexamples.

\vskip .1truein
\noindent Knowing the irreducibility of our elementary actions the algorithm to compute the configurations is as follows. First divide the bounded domain into a checkerboard of unit squares, say the black and white representations of the configuration (suppose boundary arrows are only in black). On each color in turn apply probabilistically the actions say in the sequence ${\rm I}\rightarrow{\rm II}\rightarrow{\rm III}\rightarrow{\rm IV}$ and after each action update the complementary color squares. The boundary arrows will remain fixed throughout since the boundary squares cannot be acted upon, only their (interior) arrows updated according to the arrows on the neighboring white squares. It is crucial to recognize that on a given color, the updates on its squares can be done independently everywhere (off-boundary) since the unit squares of same color share only vertices, no arrows. By definition the flips respect the 19-vertex rule i.e. vertex configurations at corners remain legal. Our initial choice of flip probabilities is as follows:


\setcounter{figure}{2}
\begin{figure}[H]
\centerline{\includegraphics[height=4cm]{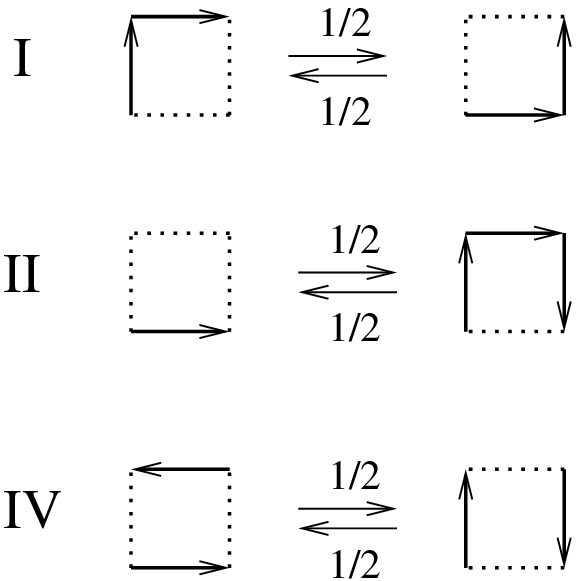}\hskip 1.5cm 
\includegraphics[height=4cm]{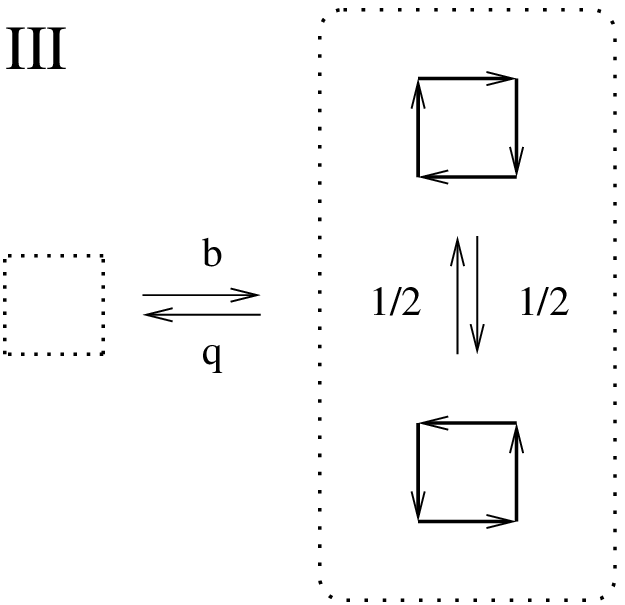}}
\caption{Flip probabilities}
\end{figure}


\noindent Once $0<b,q<1$ the Markov Chain is irreducible and aperiodic, hence ergodic.

\subsection{Boundary condition}

For the purposes of this study we exclusively use the {\bf Domain Wall Boundary Condition} (DWBC, originally due to Izergin and Korepin) on a square. The reason is two-fold: it enables comparison to Ice-model results where it is the simplest known boundary condition for generating non-trivial limit shapes. More general domain shapes and non-trivial boundary conditions exist (see e.g. \cite{E1}) but while they are useful in teasing some further details out from the model we refrain from dwelling in them here. They are essentially similar in that in them boundary segments of extremal tilt follow each other in an alternating fashion as we circumambulate a (sub)domain.

Because of our set-up the computations are actually performed on a diamond inscribing the domain square. On the diamond we have an initial condition corresponding to a NE-oriented \lq\lq ridge roof\rq\rq\ height surface (sides with tilt 0, ends with $\pm 1$ as indicated in Figure 4). This in turn enforces the DWBC on the inscribed square for all times (and the part of the diamond outside the square is frozen). 

Since all actions will be confined to the square, subsequent illustrations will be cutouts of the square alone.

While Ice configurations are fully occupied by arrows, the dynamic 19-vertex model allows density relaxation from even fully occupied initial state. By counting the minimal number of paths connecting the incoming and outgoing arrows on neighboring sides of the square one can readily deduce

\begin{prop}
In the scaling limit the arrow density of a 19-vertex configuration over a domain with DWBC is always at least $\frac{1}{2}.$
\end{prop}

\setcounter{figure}{3}
\begin{figure}[H]
\centerline{\includegraphics[height=6cm]{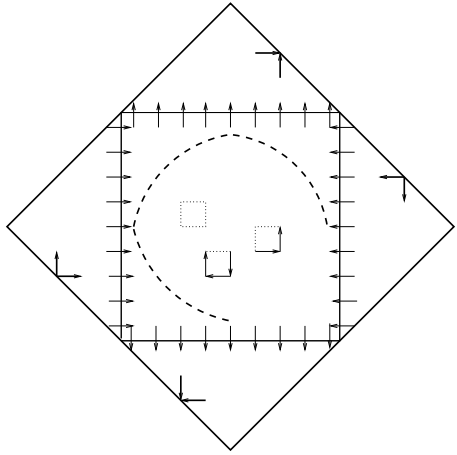}}
\caption{Domain wall boundary condition induced to the inscribed square.}
\end{figure}

\subsection{Parameter square}

Our first runs aimed to map the behavior of the model in the $(b,q)$-parameter square. The initial state was always a fully occupied (Ice-legal) configuration with DWBC (from a ridge roof height surface in the diamond). Before going in to the details of the data, a few general remarks about Figure 5.

Some things are immediate: for example everywhere at the bottom i.e. $q=0, b\in [0,1]$ the dynamics is exactly that of the Ice-model. All black/white squares are fully oriented forever and those among them actually unidirectional are randomly acted upon by a subaction of III while none of the other actions ever take place. The normalized Ice weights at the bottom are 1 hence it corresponds to the 1-enumeration point of ASMs in the disordered phase.

\setcounter{figure}{4}
\begin{figure}[H]
\centerline{\includegraphics[height=4cm]{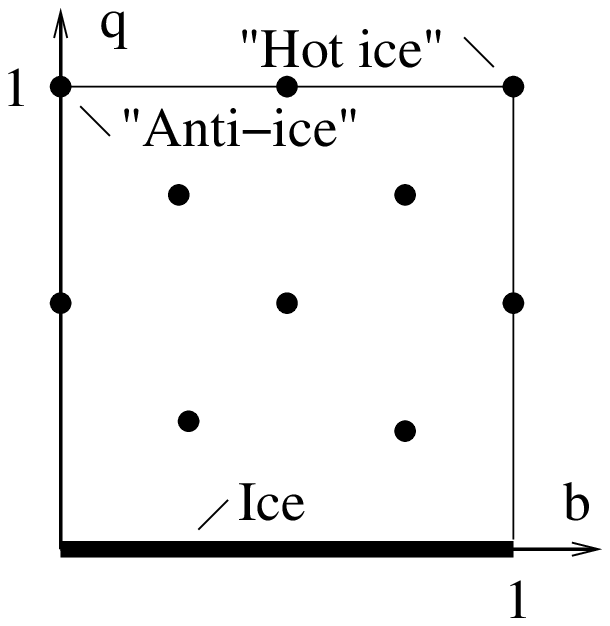}}
\caption{Parameter space and sample points.}
\end{figure}

\noindent The quantity $b+q$, the creation-annihilation rate of unidirectional 1-loops, could perhaps be called \lq\lq dynamic Ice-temperature\rq\rq, since it indicates how high the intensity is around the unidirectional 1-cycles that generate the Ice-dynamics (once $q>0$). At the point $(1,1)$ this dynamics is fastest yet this does not materialize in the equilibrium. While on the diagonal $q=b,\ b>0$ the speed of the dynamics varies, the equilibria are all equal since the (static) weights are all 1.

The point $b=0,\ q=1$ will be of paramount interest. Under this dynamics whenever a unidirectional 1-loop appears (as a consequence of actions I, II or IV in adjacent squares), it is immediately killed by a subaction of III. Moreover no unidirectional loops are ever born out of fully unoriented 1-loops. Hence this dynamics has no Ice-component (unidirectional 1-loop reversals) in it and we will refer to it as {\bf Anti-ice}. Similar behavior of prevails (somewhat less cleanly) all along the top of the diagram: with $b>0$ the unidirectional 1-loops do keep appearing but they are transient, all killed in one iterate.

\section{Simulation results}

In order to map the bounded 19-vertex model landscape we simulated the dynamics at 10 uniformly distributed points in the parameter space (the black dots in Figure 5) and at one more at the bottom. The size of the diamond was $212\times 212$ resulting in $150\times 150$ square domain on which the DWBC was in effect.

At every point of our parameter square the dynamics resulted in a distinct limit shape. This was plain from observing the configuration evolutions alone. The shape typically stabilized well before 20-30.000 iterates through both black and white lattices. During this relaxation period no data was gathered. Afterwards a further run of length at least 20.000 iterates was performed to get the equilibrium distributions of the elementary actions. They are essentially the cumulative flip counts for each type of action (subactions for III) at every site in the square. Next we present some of the key details screened from this action data.

\subsection{Diagonal $q=1-b$}

In Figure 6 the bottom row shows the sole Ice action distribution over the square. What is in fact rendered is the cumulative count of unidirectional 1-cycles over the domain (half of which are flipped under the dynamics). The shape of the distribution boundary is the limit shape which is known to corresponds to the 1-weight ASM (in the disordered regime). It is close but not quite circular (\cite{CP}).

The middle row corresponds to the center point of the parameter space, but as indicated earlier, this set of equilibrium distributions is shared by all the points on the rising diagonal $q=b,\ b>0.$ Here first subplot of III shows the cumulative count of fully unoriented 1-cycles at each site. Under the dynamics fraction $b$ of them will be turned unidirectional (with both orientations equally likely) in one iterate. The second subplot of III shows the cumulative distribution of unidirectional 1-cycles; in one step fraction $q$ of these will be annihilated, half of the rest reversed, the other half left intact.

\setcounter{figure}{5}
\begin{figure}[H]
\footnotesize
\centering
\begin{tabular}{ccccc}
 \medskip
\stackunder{\includegraphics[width = 1.05in]{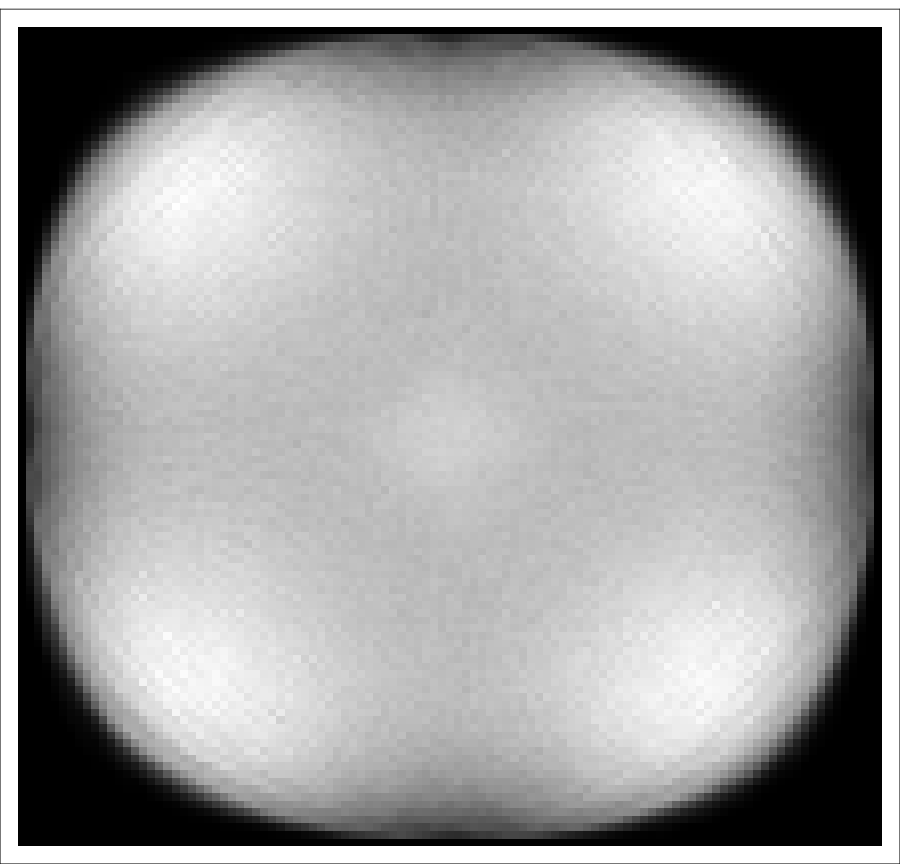}}{I} &
\stackunder{\includegraphics[width = 1.05in]{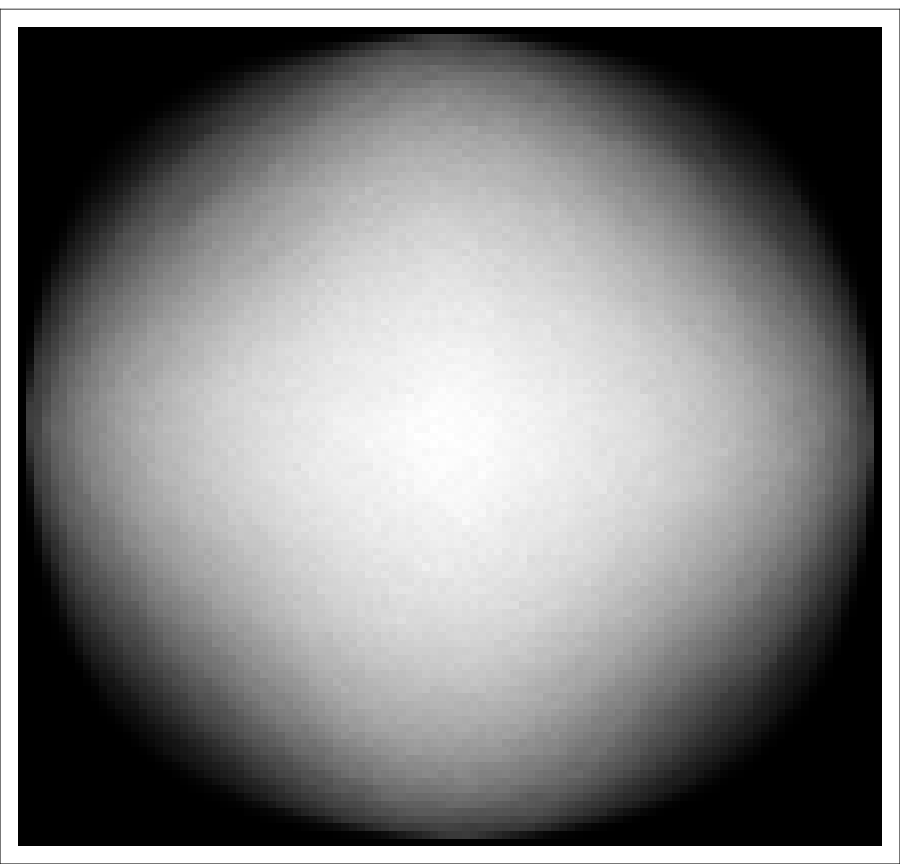}}{II} &
\stackunder{\includegraphics[width = 1.05in]{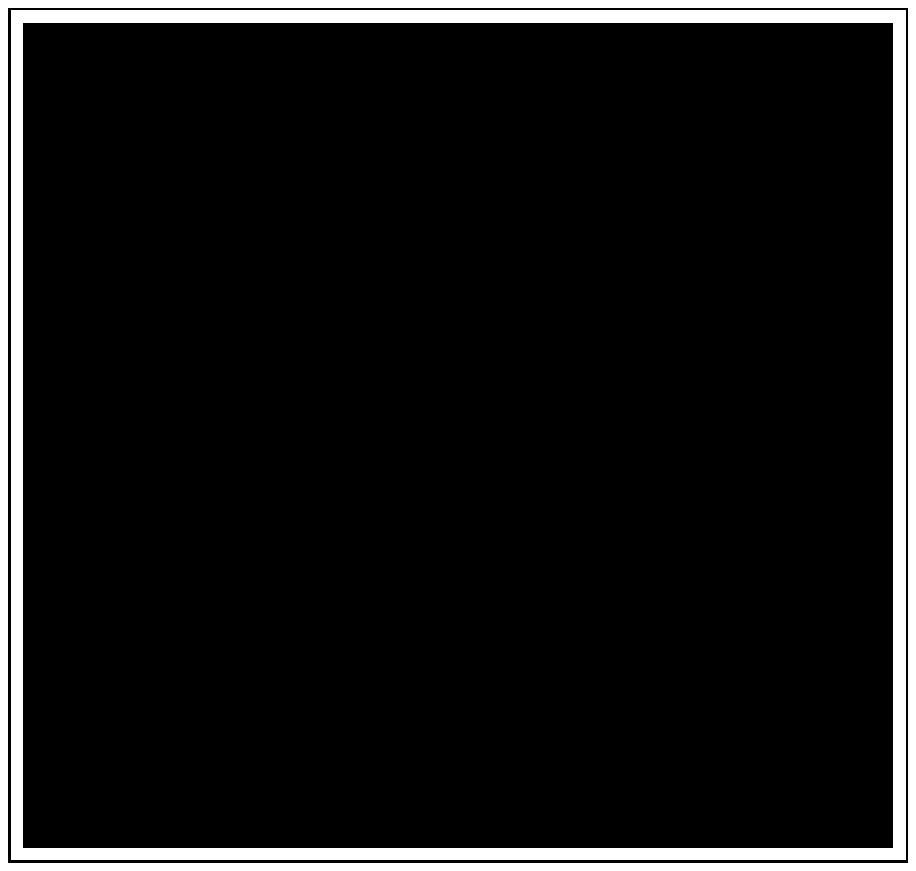}}{III, births} &
\stackunder{\includegraphics[width = 1.05in]{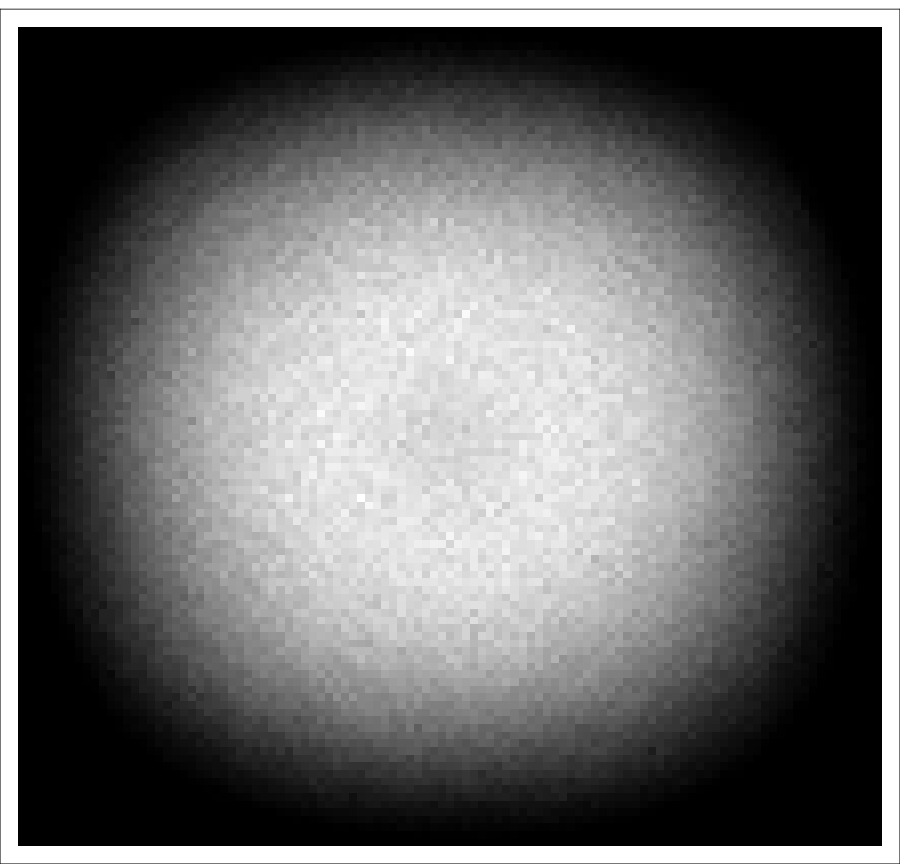}}{III, annihilations} &
\stackunder{\includegraphics[width = 1.05in]{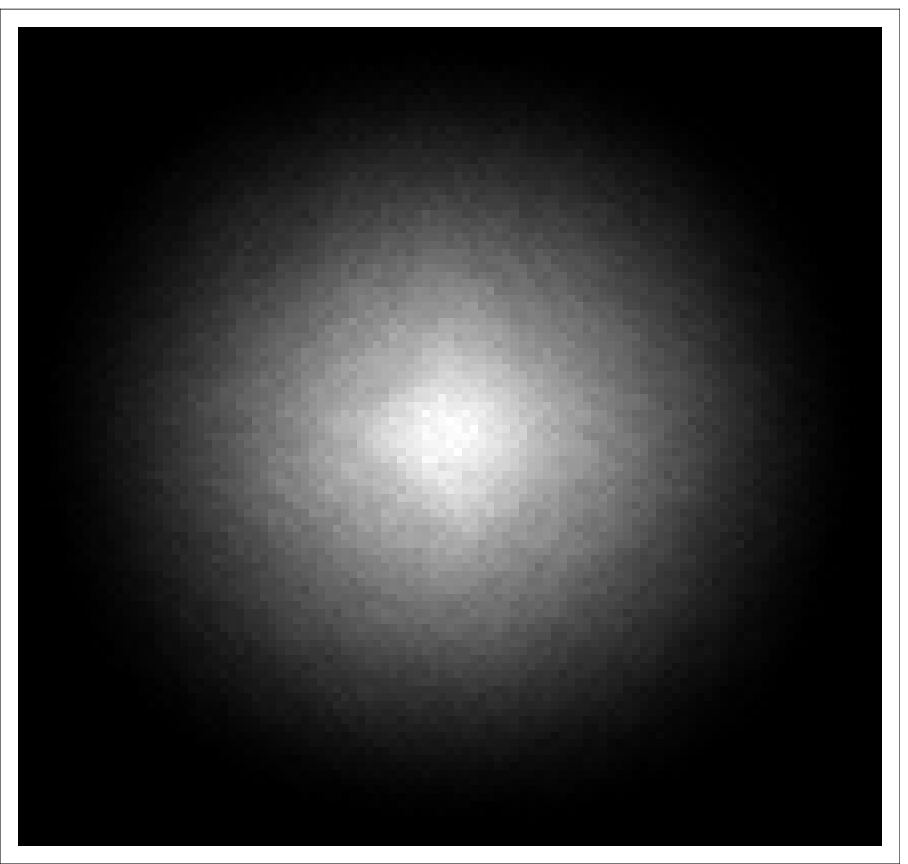}}{IV}\\
 \medskip
\stackunder{\includegraphics[width = 1.05in]{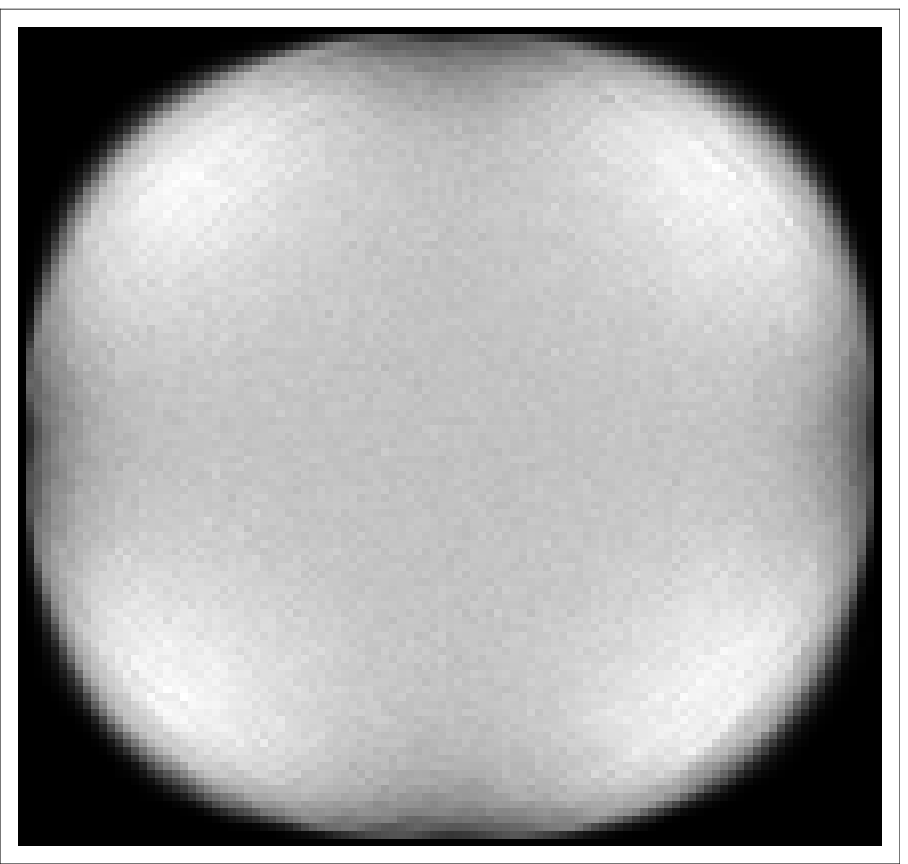}}{I} &
\stackunder{\includegraphics[width = 1.05in]{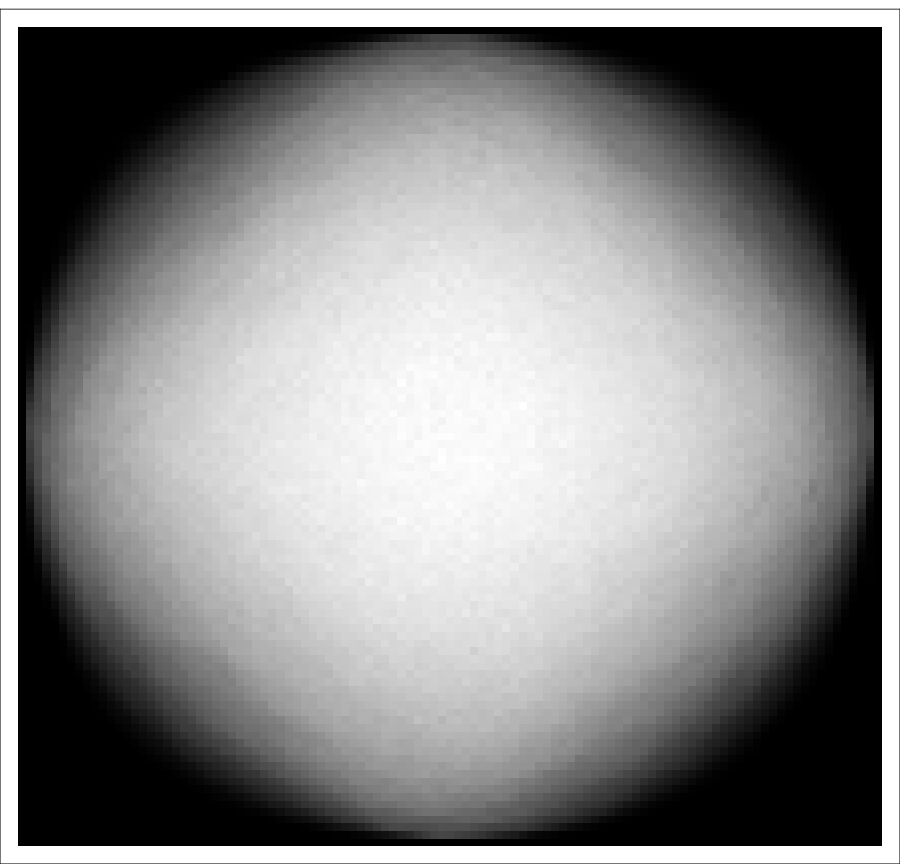}}{II} &
\stackunder{\includegraphics[width = 1.05in]{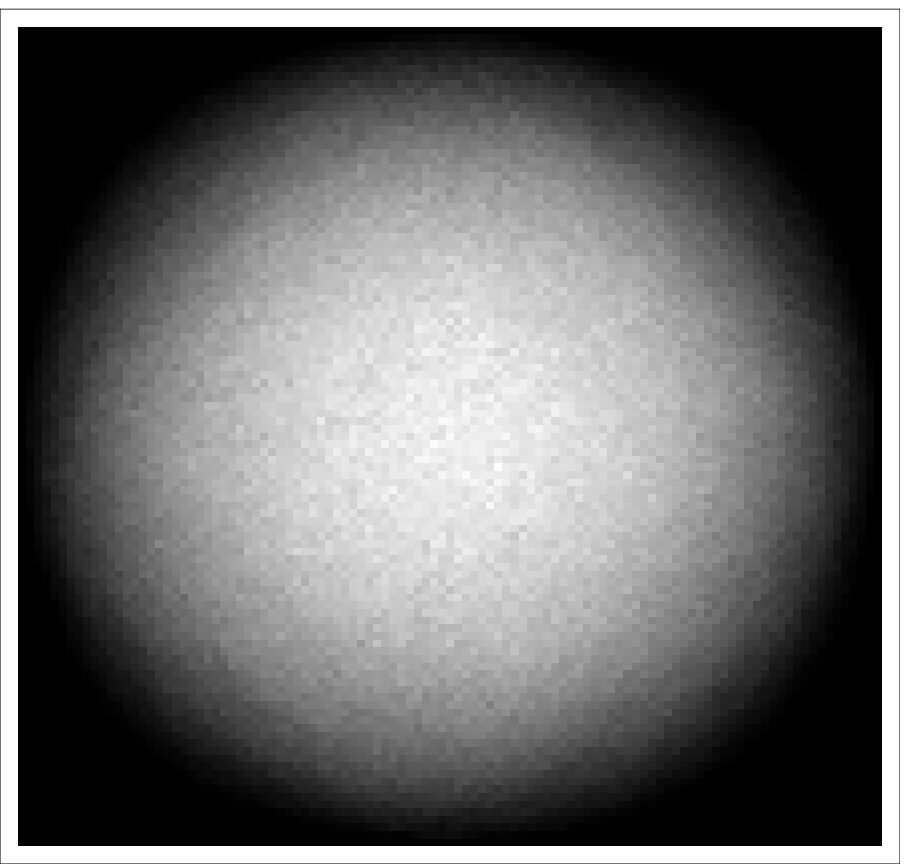}}{III, births} &
\stackunder{\includegraphics[width = 1.05in]{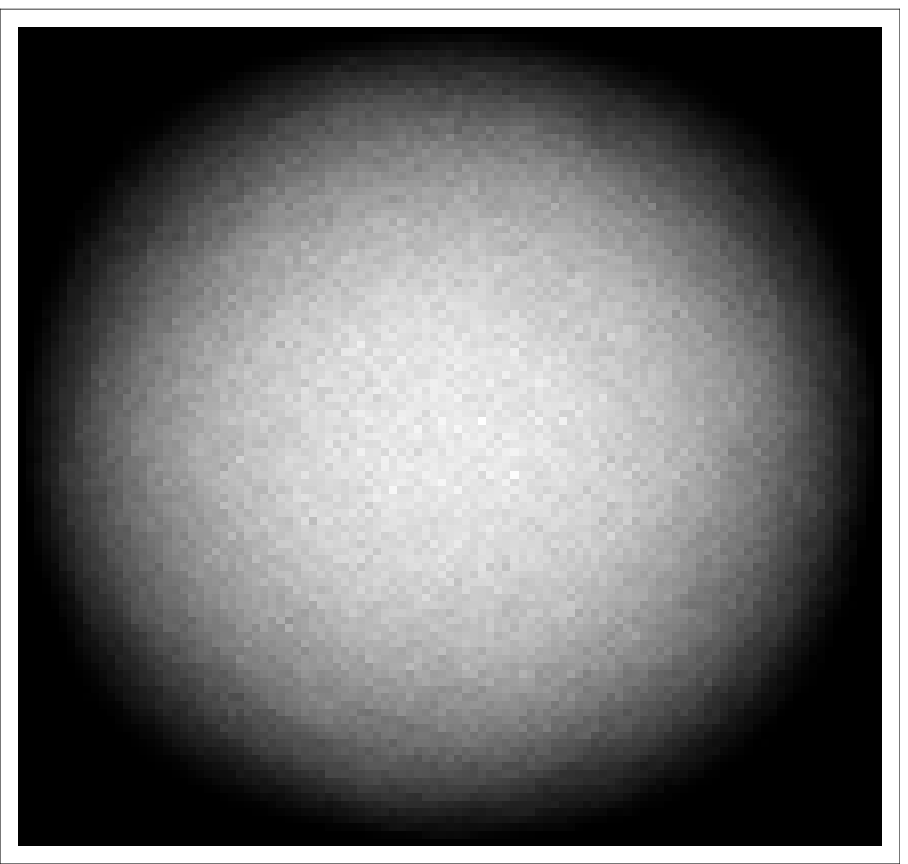}}{III, annih. or revers.} &
\stackunder{\includegraphics[width = 1.05in]{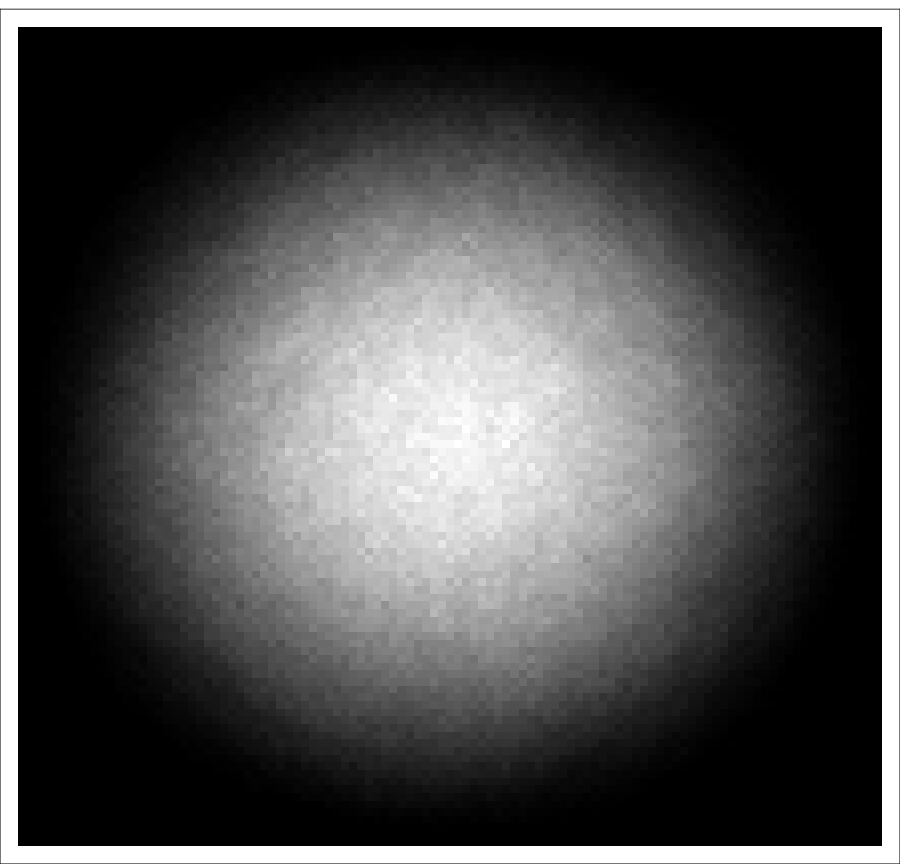}}{IV}\\
 \medskip
\stackunder{\includegraphics[width = 1.05in]{a-black.eps}}{I} &
\stackunder{\includegraphics[width = 1.05in]{a-black.eps}}{II} &
\stackunder{\includegraphics[width = 1.05in]{a-black.eps}}{III, births} &
\stackunder{\includegraphics[width = 1.05in]{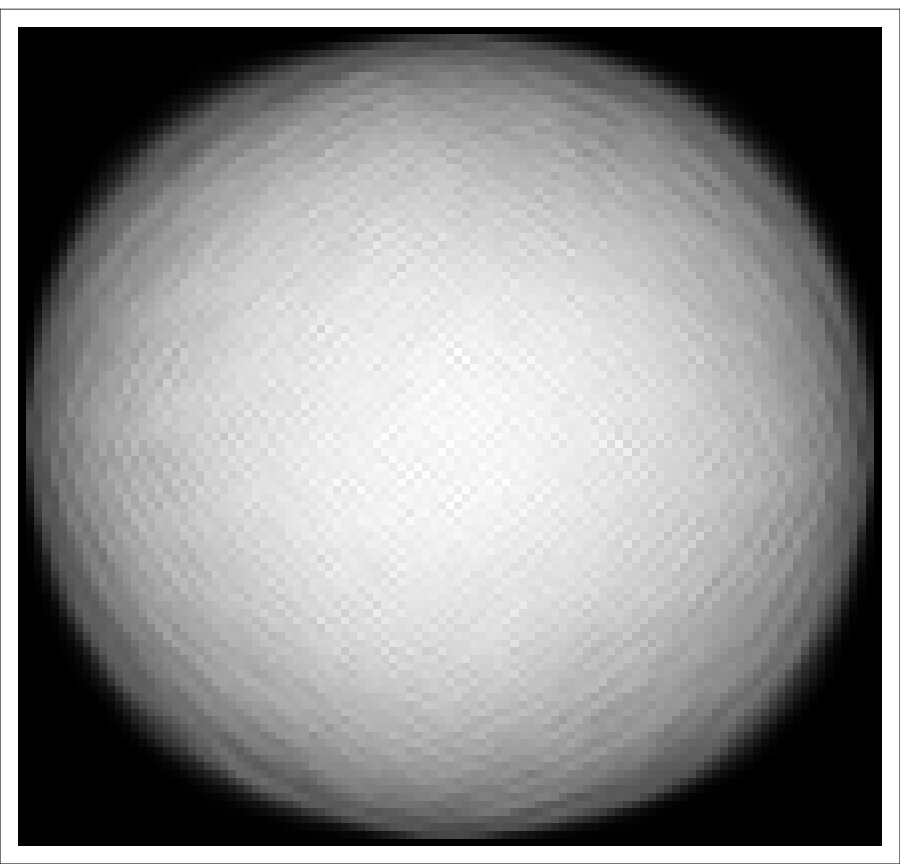}}{III, reversals} &
\stackunder{\includegraphics[width = 1.05in]{a-black.eps}}{IV}\\
\end{tabular}

 \caption{Action densities. The lighter the more active while black means no activity. (i) bottom row: 1-weight Ice ($(b,0),$ any $b$, (ii) middle row: diagonal $q=b>0$ and (iii) top row: Anti-ice ($(b,q)=(0,1)$). Captions indicate the applicable (sub)actions.}

\end{figure}

\noindent 
The top row lays out the Anti-ice densities. No births or reversals take place under the type III action here, hence the subaction is annihilation alone. That empirical density as well as the three others are subtly different from those of the center row. What is also clear is that the limit shape will be forming in spite of the missing Ice-dynamics. This can readily seen from the configurations, too, but here as well as elsewhere in our study a more informative picture emerges from action distributions and this is why we prefer to analyze them.

A rather prominent feature in the two top rows as well as in similar data from all of our other data points (off-Ice i.e. when $q>0$) is that in 19-vertex model action I dominates in determining the limit shape. The support of its density seems to contain the others and the density also has the steepest sides resulting in sharpest boundary. This is a signature of a spatial phase transition. For another view of this, in Figure 7 we have plotted the horizontal and diagonal sections of the Anti-ice densities. Only even sublattice data is rendered, the odd sublattice looks essentially the same.

\setcounter{figure}{6}
\begin{figure}[H]
\footnotesize
\centering
\begin{tabular}{cccc}
 \bigskip
\stackunder[5pt]{\includegraphics[width = 1.3in]{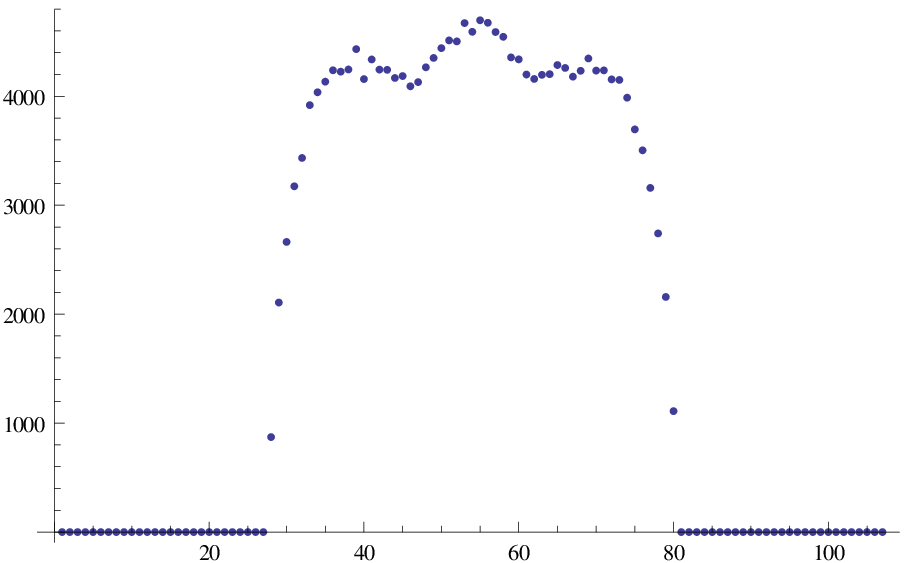}}{I} &
\stackunder[5pt]{\includegraphics[width = 1.3in]{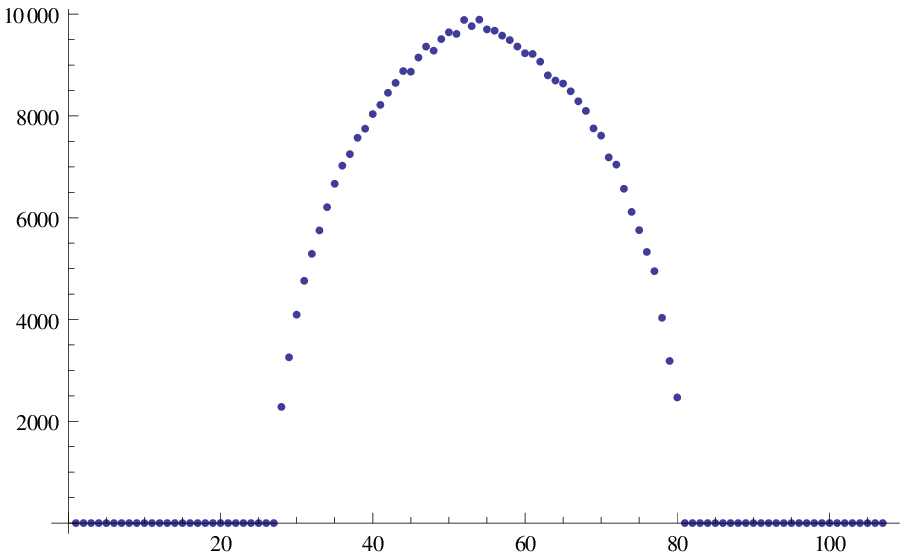}}{II} &
\stackunder[5pt]{\includegraphics[width = 1.3in]{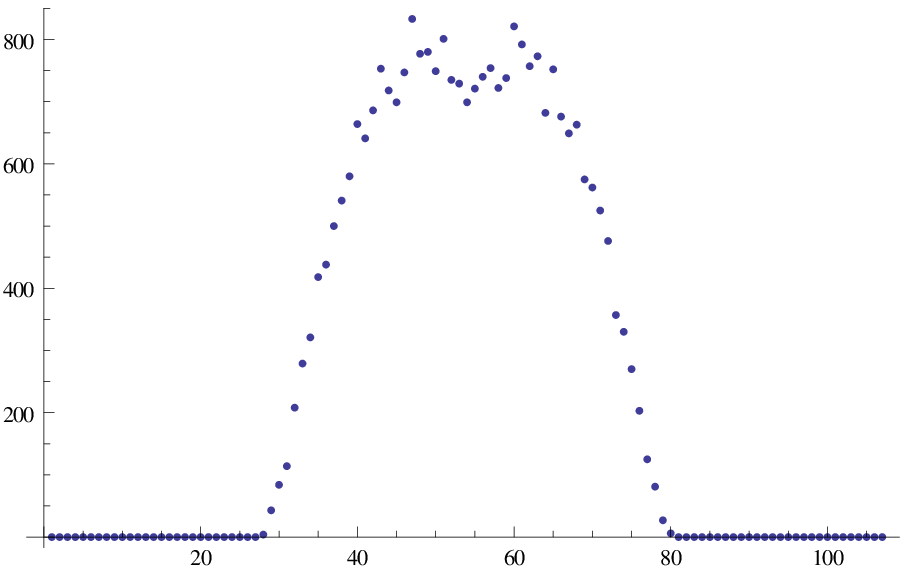}}{III, annihilations} &
\stackunder[5pt]{\includegraphics[width = 1.3in]{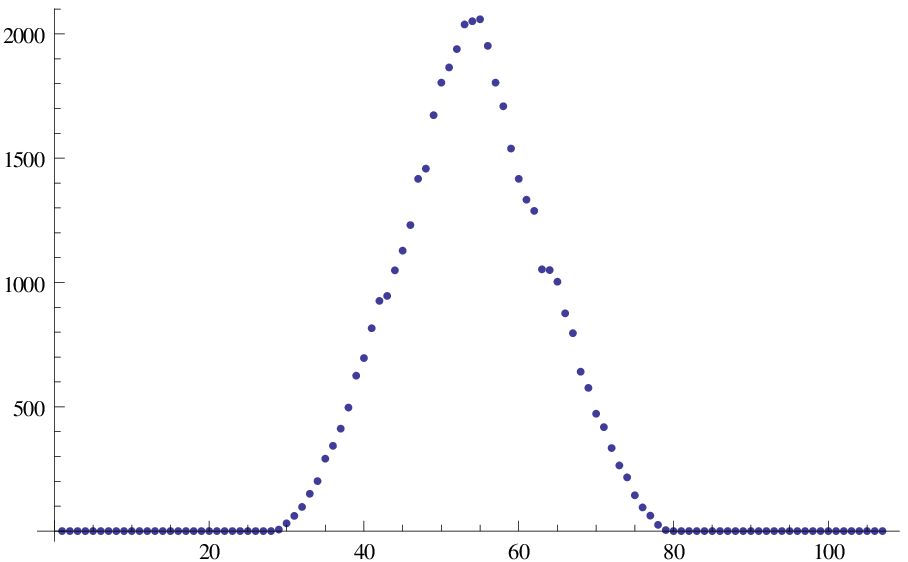}}{IV}\\
 \bigskip
\stackunder[5pt]{\includegraphics[width = 1.3in]{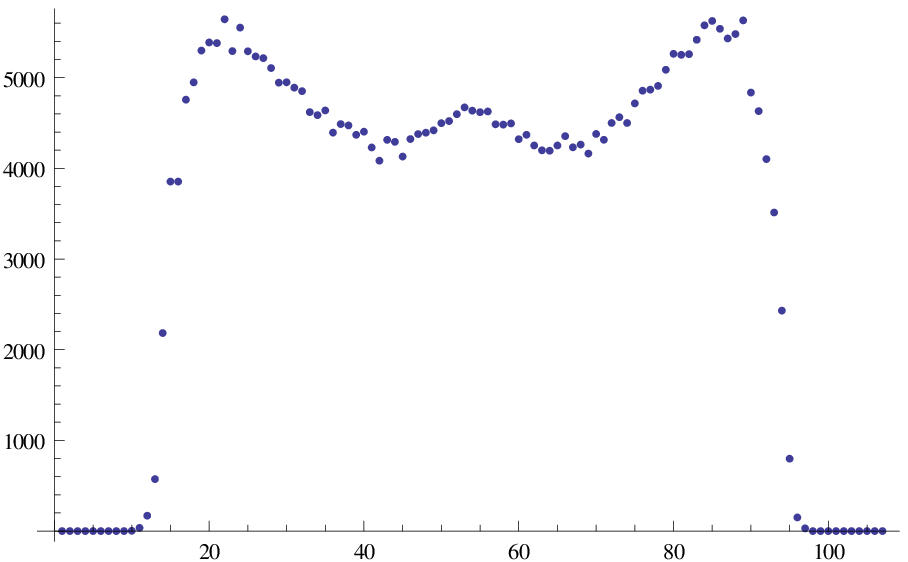}}{I} &
\stackunder[5pt]{\includegraphics[width = 1.3in]{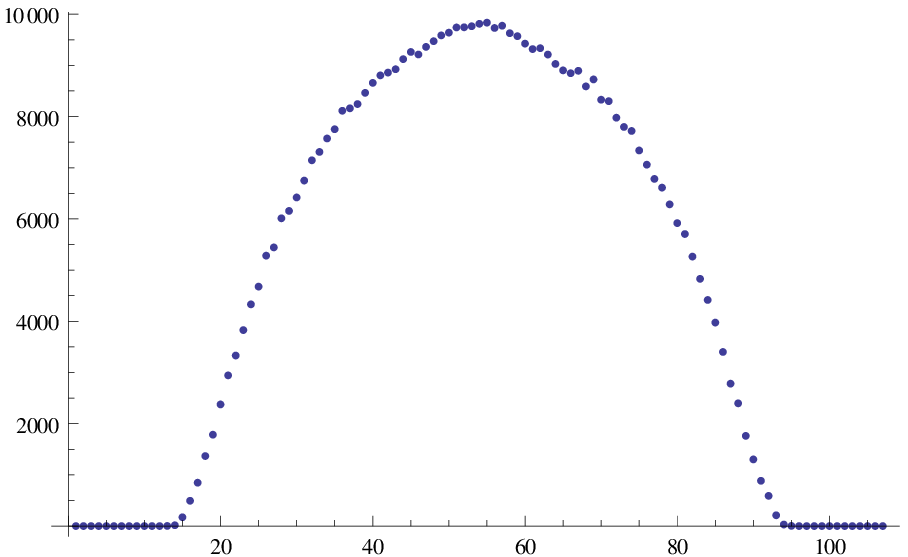}}{II} &
\stackunder[5pt]{\includegraphics[width = 1.3in]{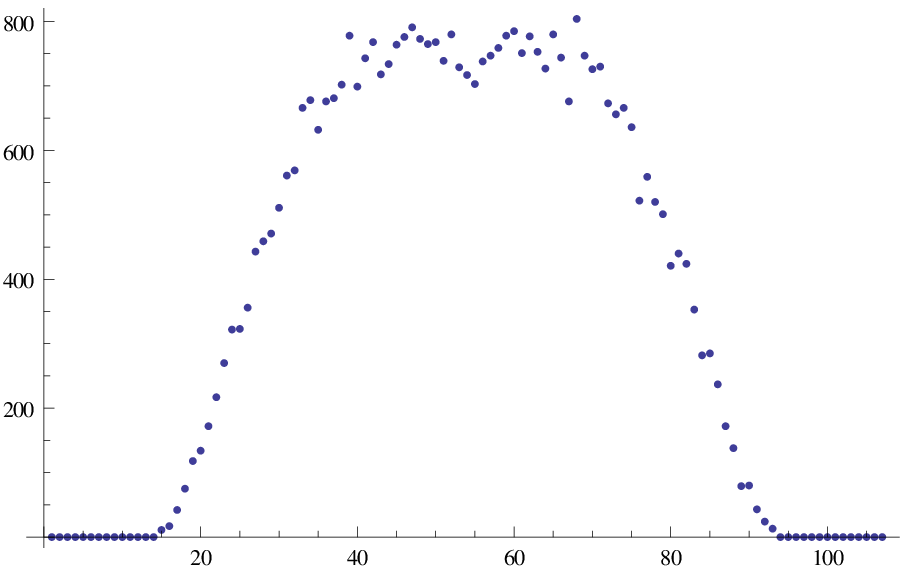}}{III, annihilations} &
\stackunder[5pt]{\includegraphics[width = 1.3in]{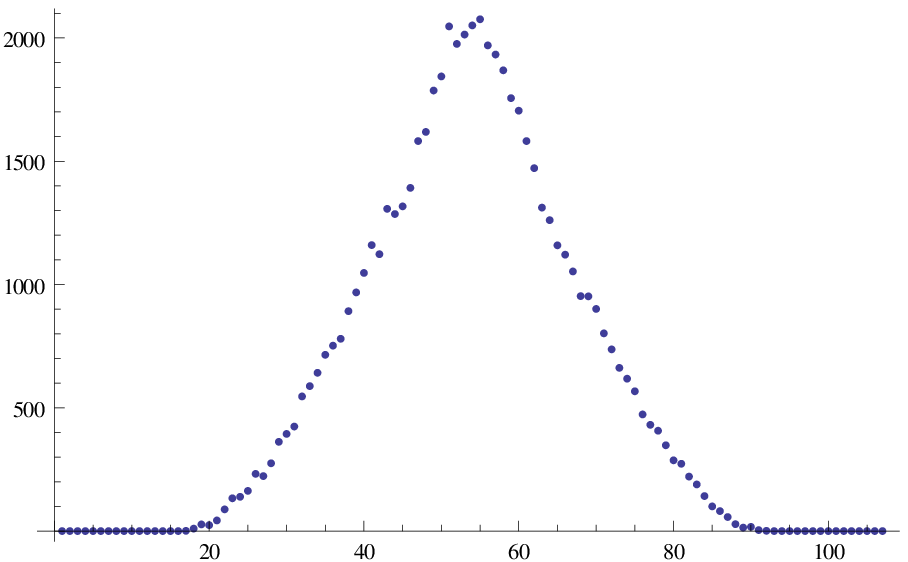}}{IV}\\
\end{tabular}

\caption{Anti-ice action density sections: horizontal (top row) and diagonal (bottom row) cut through the center of the domain. These are cuts from the entire diamond hence the longer abscissae.}

\end{figure}

\noindent The action I sections clearly show the concentration of activity in the diagonal directions as one should expect given the orientation of the underlying lattice. We do not have an explanation for the center hump but for the limit shape this may be of secondary interest. The same is likely to be true regarding action IV, too. Indeed the relative volumes of the distributions may be relevant. In Figures 6 and 7 the vertical scales are chosen individually for the types for best rendering.

The maxima of the action densities (combined from black and white) over the square are illustrated in Figure 8. On the rising diagonal $q=b$ of the parameter square we have just five sample points (uniformly on $[0,1]$) hence the somewhat jagged look. For $b\ge\frac{1}{4}$ the graphs should in fact be constant - the slight variation from horizontal are due to random fluctuations in the data. 

\setcounter{figure}{7}
\begin{figure}[H]
\centerline{\includegraphics[height=4cm]{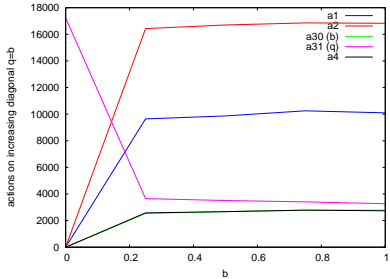}}
\caption{Action maxima at sample points along $q=b$.}
\end{figure}

\noindent Once all the actions are at possible ($q=b>0$) the highest volume action is II followed by I with less than 60\% of the intensity, while the other actions are much less frequent. Whether action II in fact \lq\lq feeds\rq\rq\ the others in an essential way is an interesting question (near the boundaries of the parameter square this can certainly be argued). The large gaps in intensities seems to further underline the insignificance of Ice action (reversals in III) in the 19-vertex dynamics.

\subsection{Skewed Action II}

\noindent Our algorithm can be modified in a multitude of ways. One could put relative weights on the entire set of actions I-IV or introduce further asymmetry within the actions like we have done with type III. Some dynamic weight combinations are bound to be more interesting than others as we have seen e.g. on the diagonal of Figure 5. Over weighting action I would likely not change the limit shape much at all and the same should be true for IV, too, since it seems to be active principally in the central region.

Of our actions only II and III (the two non-Ice subactions) change the arrow density locally. In Anti-ice we saw some of the consequences of this kind of quenching in III ($q=1$ annihilating all arrows in unidirectional 1-cycles). We will now indicate how this plays out with action II.

\setcounter{figure}{8}
\begin{figure}[H]
\centerline{\includegraphics[height=1cm]{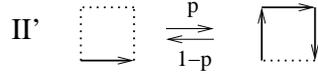}}
\caption{Action II parametrized.}
\end{figure}

\noindent One can readily grasp the microscopic effect of this modification (applied to reflections and rotations as well), its local influence on oriented paths. As $p$ decreases paths straighten out, while $p$ increasing they convolute further. At the extreme $p\approx 0$ one should see subdomains with sparse collections of short, oriented paths and therefore limited action on them. Conversely $p\approx 1$ should result in ensembles of highly meandering, even locally space-filling directed curves. This should in turn provide sites for the other flip types to kick in. The weighting in II' does not result in total local arrow annihilation/birth out of vacuum (as in action III) but rather a geometric effect which in turn influences the arrow density.

In order to see what this might imply macroscopically we gathered some data around a few of our sample points in the $(b,q)$ -parameter square by additionally varying also the new parameter $p.$ The size of the domain, initial state, the simulation strategy and the lengths of the runs were essentially as before. Again we only show equilibrium action densities instead of configurations snapshots. Here are results from two such perturbations of Anti-ice:

\setcounter{figure}{9}
\begin{figure}[H]
\footnotesize
\centering
\begin{tabular}{cccc}
 \medskip
\stackunder{\includegraphics[width = 1.35in]{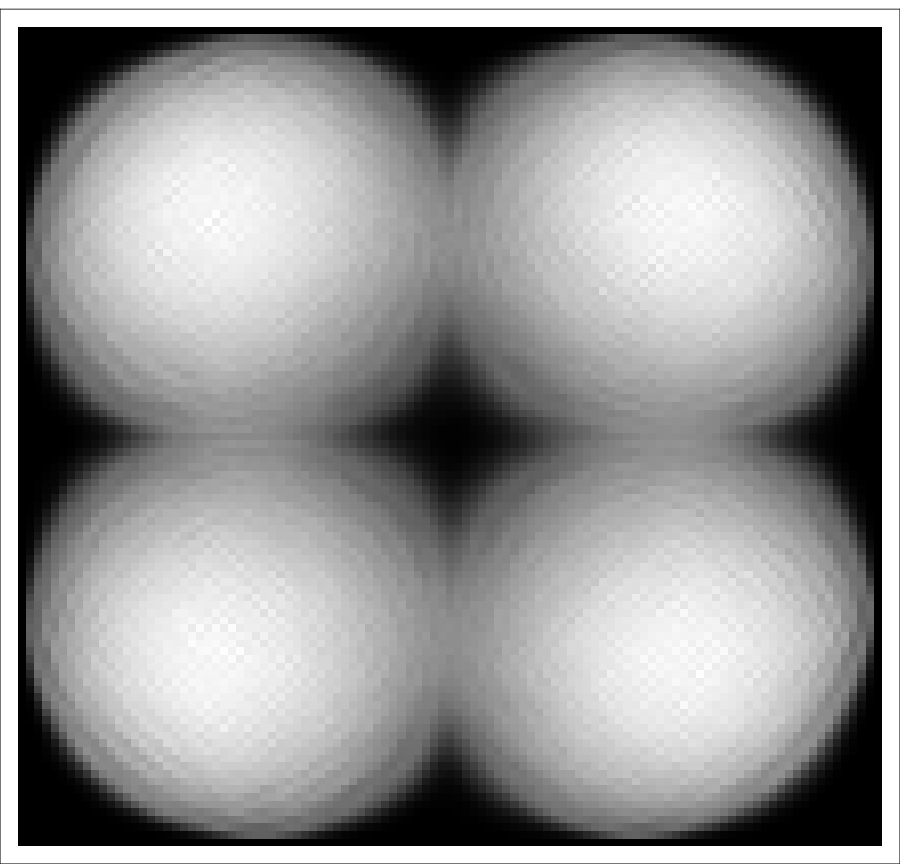}}{I} &
\stackunder{\includegraphics[width = 1.35in]{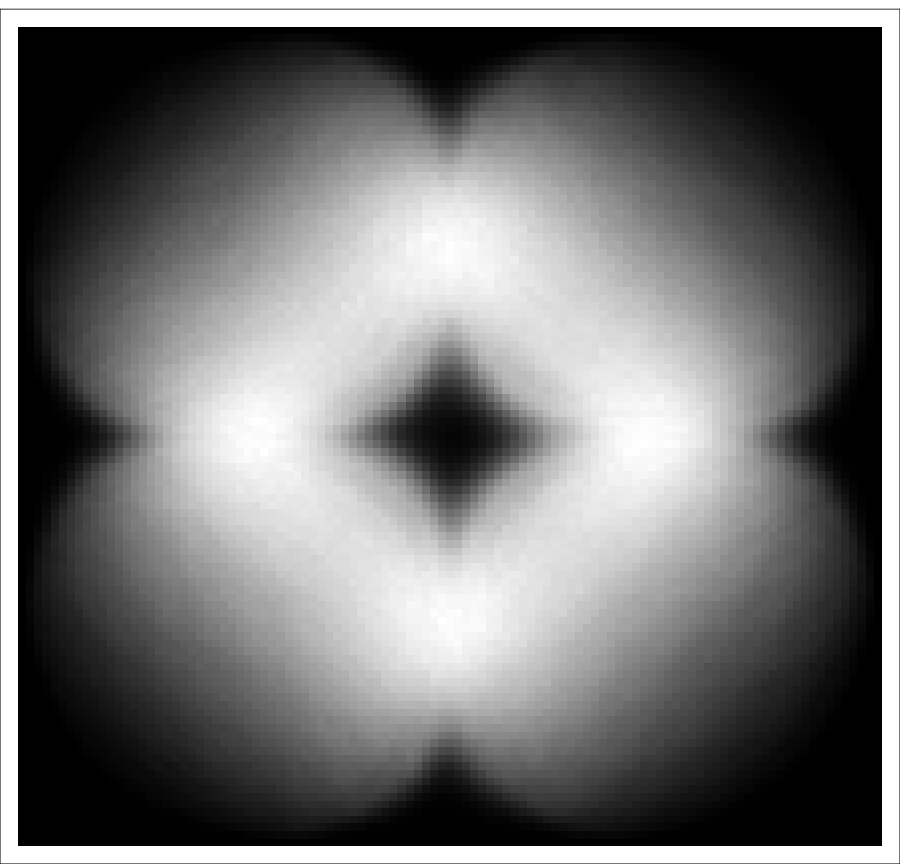}}{II'} &
\stackunder{\includegraphics[width = 1.35in]{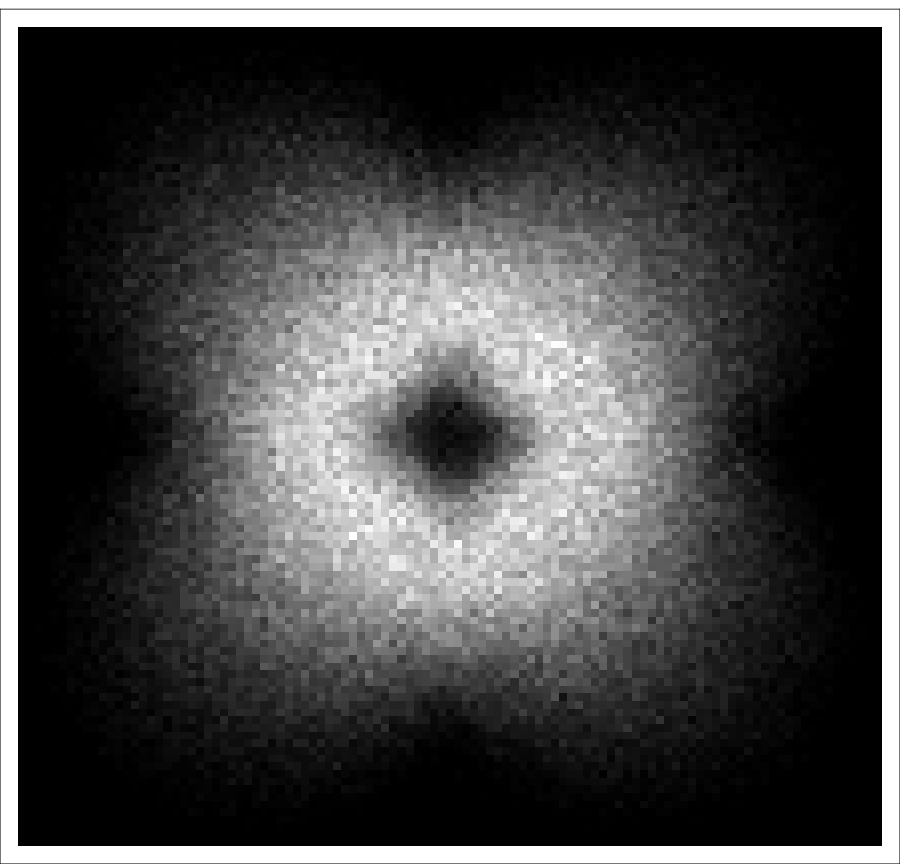}}{III, annihilations} &
\stackunder{\includegraphics[width = 1.35in]{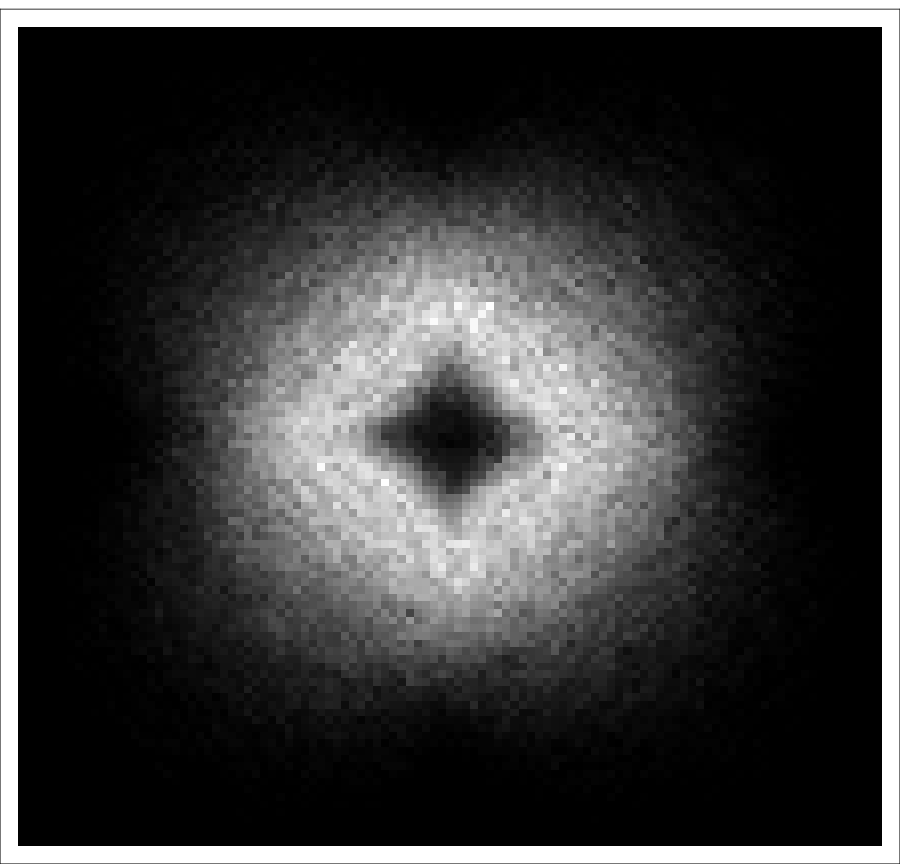}}{IV}\\
 \medskip
\stackunder{\includegraphics[width = 1.35in]{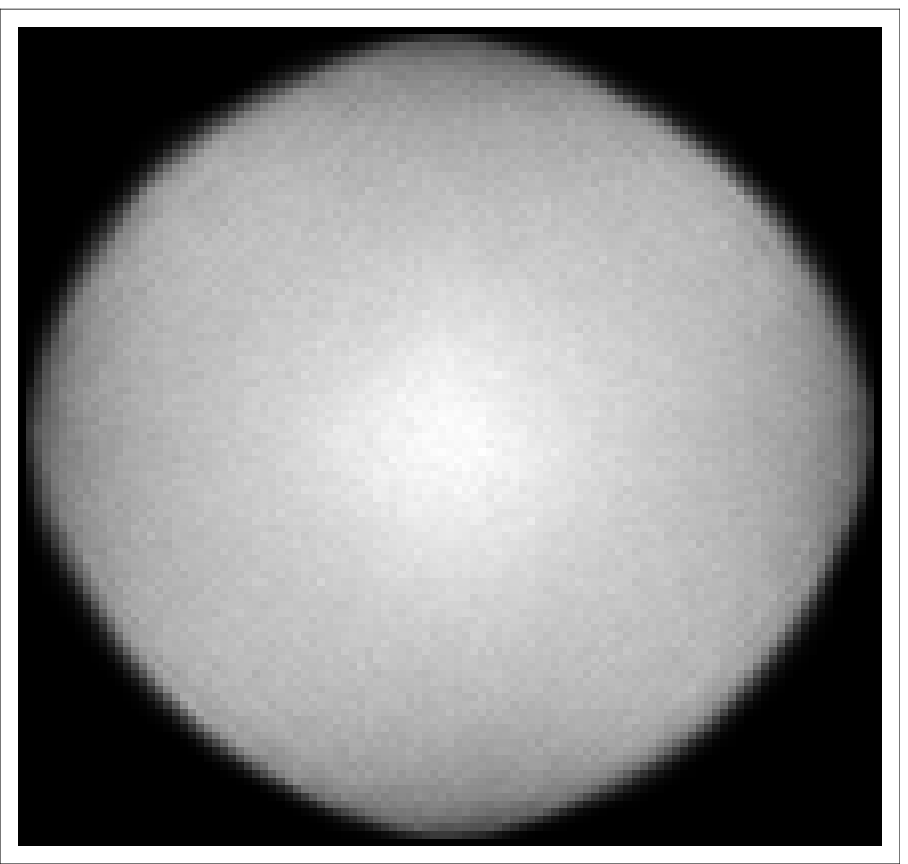}}{I} &
\stackunder{\includegraphics[width = 1.35in]{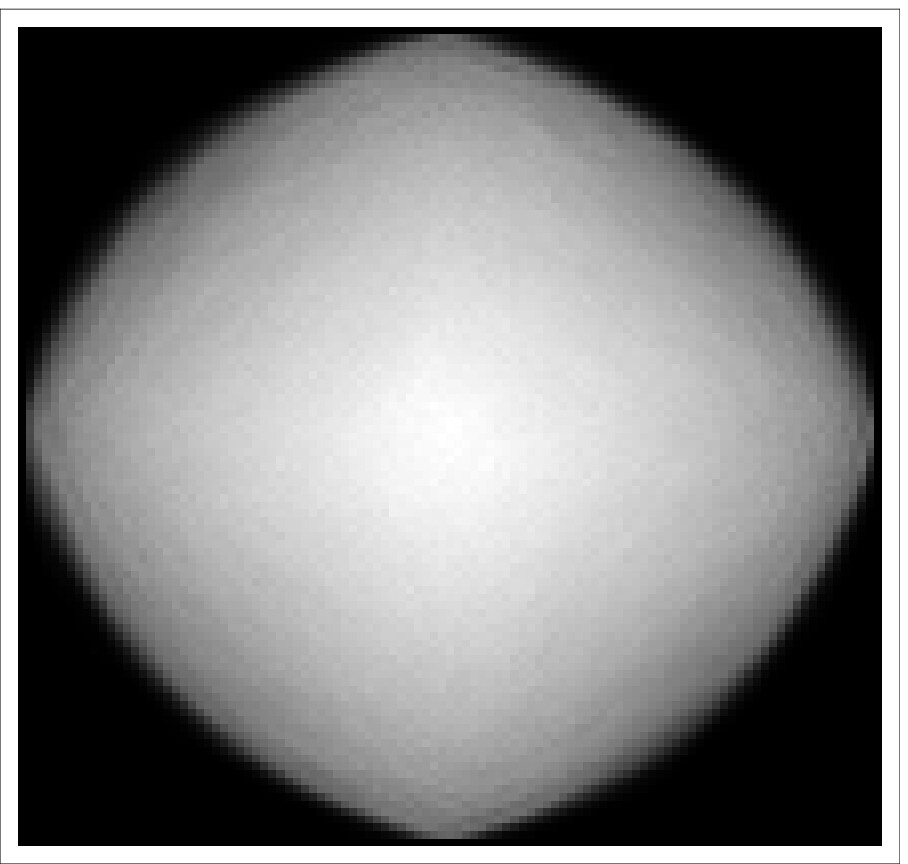}}{II'} &
\stackunder{\includegraphics[width = 1.35in]{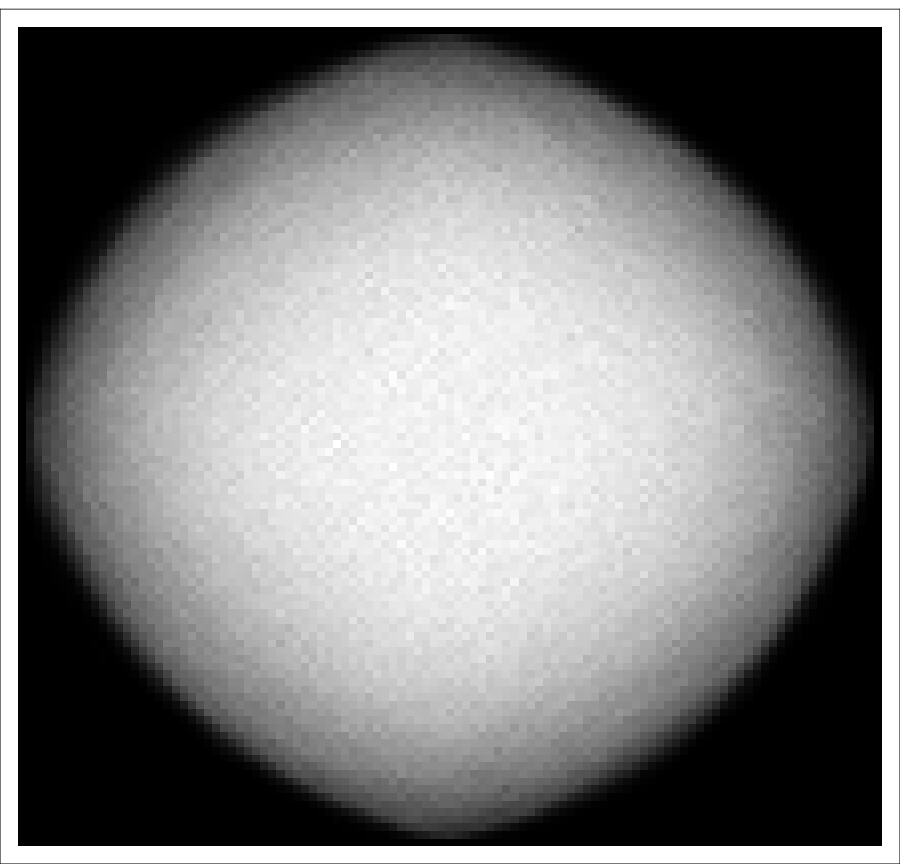}}{III, annihilations} &
\stackunder{\includegraphics[width = 1.35in]{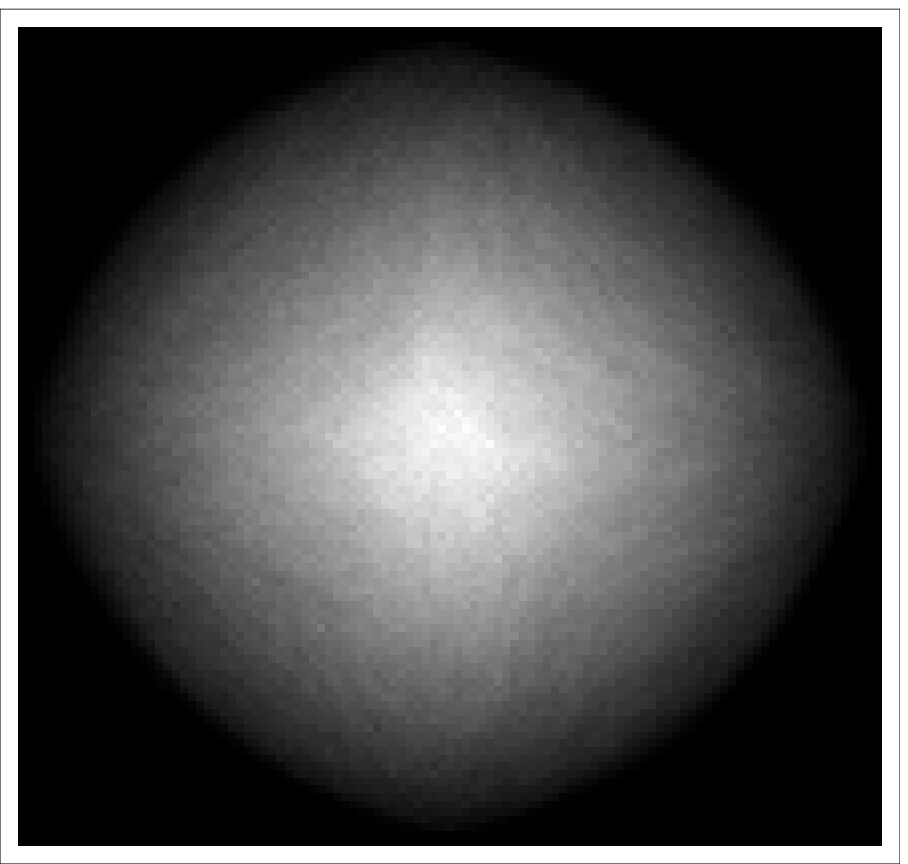}}{IV}\\
\end{tabular}

 \caption{Densities at $(b,q)=(0,1)$ under $p$-variation. (i) top row: $p=\frac{1}{8}$, and (ii) bottom row: $p=\frac{7}{8}.$ Columns indicate the applicable actions, I-IV. Between these two rows one could insert the top row of Figure 6 (without the blank square), since Anti-ice corresponds to the balanced case $p=\frac{1}{2}.$}

\end{figure}

\noindent These runs seem to confirm the intuition on the local behavior of the paths. Quenching at and below $p=\frac{1}{8}$ drives the arrow density so low at the center of the domain that little action takes place. From the illustrations above it might seem like an interior boundary is born and that it might even be shared by all the action distributions. This is likely not the case. Firstly the distributions III and IV are almost two orders of magnitude thinner than the two others so no firm conclusions should be drawn from their data. Secondly while at $p=\frac{1}{4}$ the local density minima of all of the distributions at the center are positive, the distribution sections at $p=\frac{1}{8}$ (below) indicate that the curves just kiss the horizontal axis at the center. This remains the case for e.g. $p=\frac{1}{16}$ i.e. no macroscopic empty sea forms at the center for any of the actions.

Once $b>0$ the sparse center starts to fill with unidirectional 1-cycles popping up and inducing other actions. The (exterior) boundary shape seems to remain essentially the same as $b$ increases.

The case $p=\frac{7}{8}$ is strikingly different from its complementary case $p=\frac{1}{8}$ as well as from Anti-ice. All the action densities seem to have roughly the same kind of support and their intensities are now in the same order of magnitude. Distributions I-III may all lead to the same limit shape while the distribution IV indicates a qualitatively different, essentially linear decay towards the boundary and hence somewhat obscures its possibly distinct limit shape. Again from the distribution sections one can see that the densities are far from identical in the interior.

Other samples e.g. points $(b,q)=(\frac{1}{2},1),\ (1,1)$ and $(1,\frac{1}{2})$ seem to conform with the findings above for the complementary $p$ values: without exceptions limit shapes of distinctly different character show up for 19-vertex model at each point.

\setcounter{figure}{10}
\begin{figure}[H]
\footnotesize
\centering
\begin{tabular}{cccc}
 \bigskip
\stackunder[5pt]{\includegraphics[width = 1.3in]{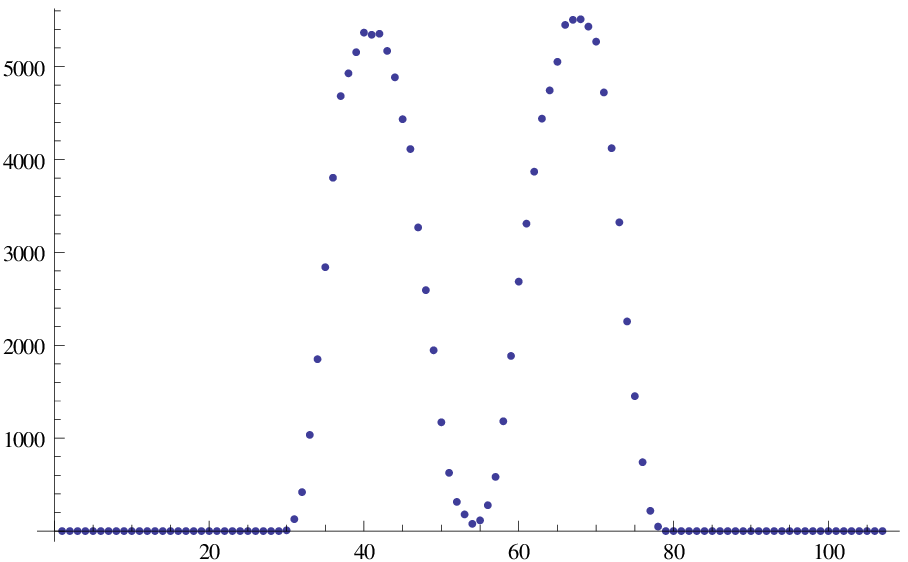}}{I} &
\stackunder[5pt]{\includegraphics[width = 1.3in]{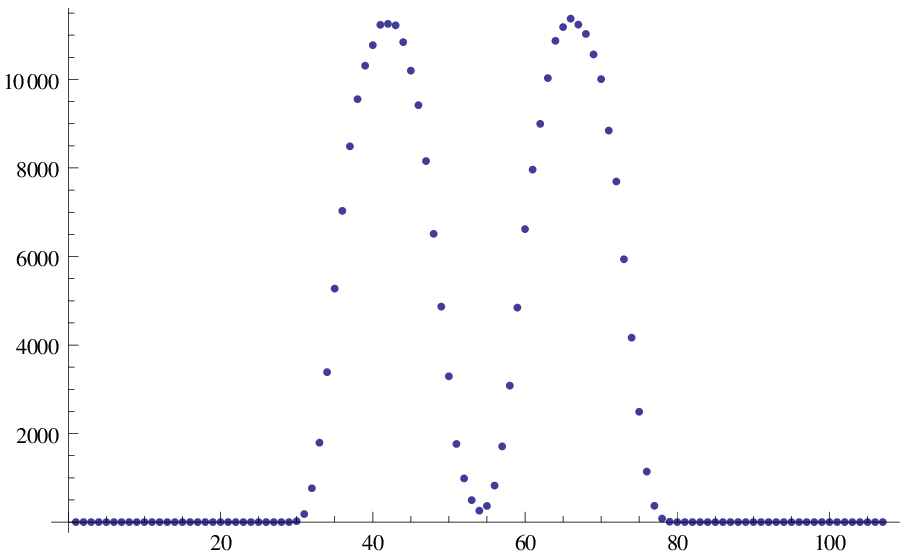}}{II'} &
\stackunder[5pt]{\includegraphics[width = 1.3in]{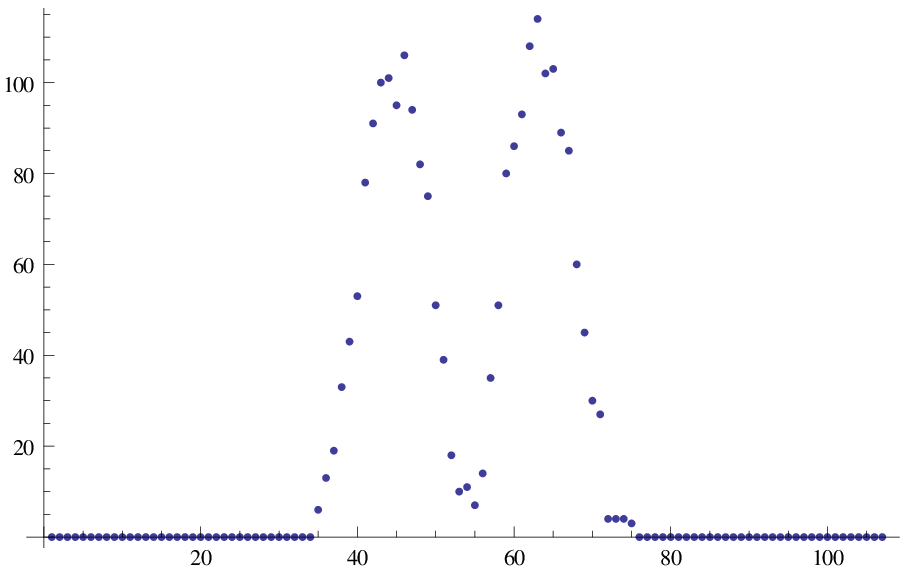}}{III, annihilations} &
\stackunder[5pt]{\includegraphics[width = 1.3in]{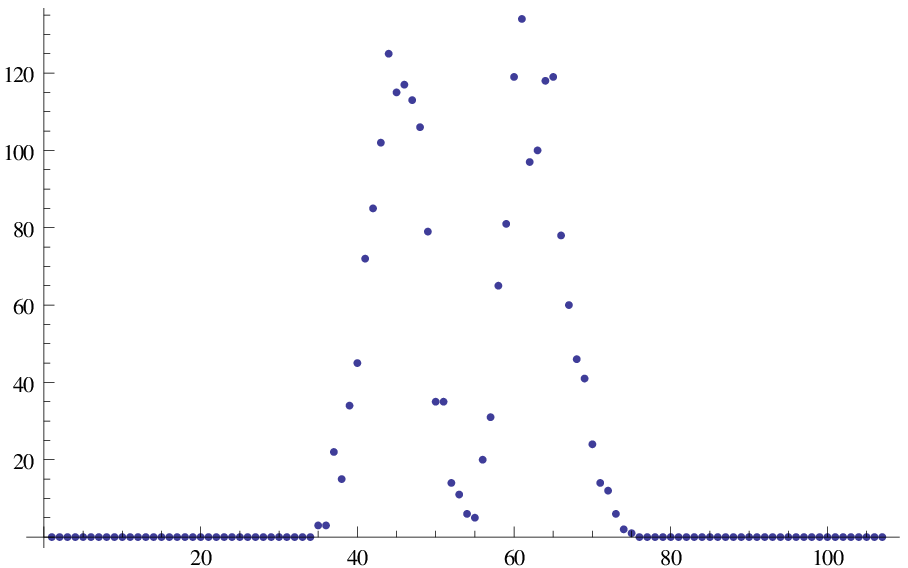}}{IV}\\
 \bigskip
\stackunder[5pt]{\includegraphics[width = 1.3in]{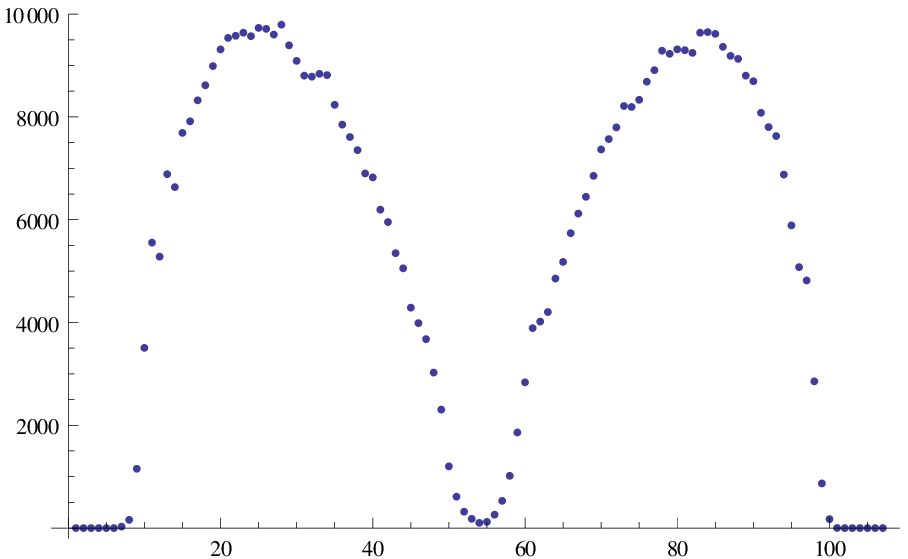}}{I} &
\stackunder[5pt]{\includegraphics[width = 1.3in]{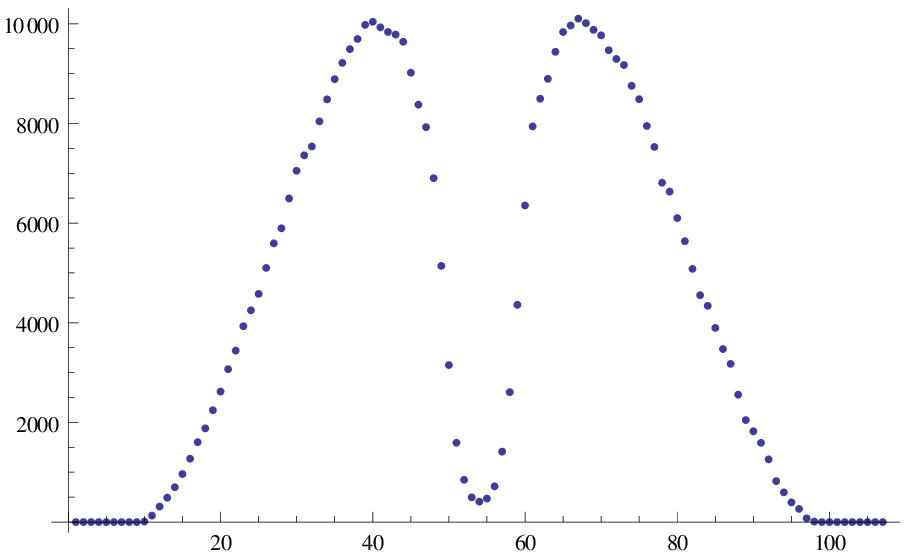}}{II'} &
\stackunder[5pt]{\includegraphics[width = 1.3in]{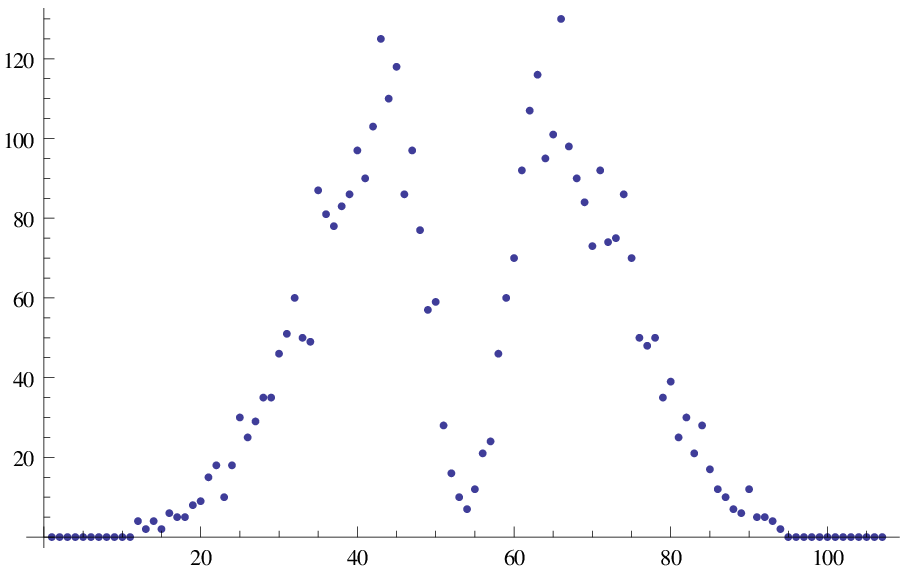}}{III, annihilations} &
\stackunder[5pt]{\includegraphics[width = 1.3in]{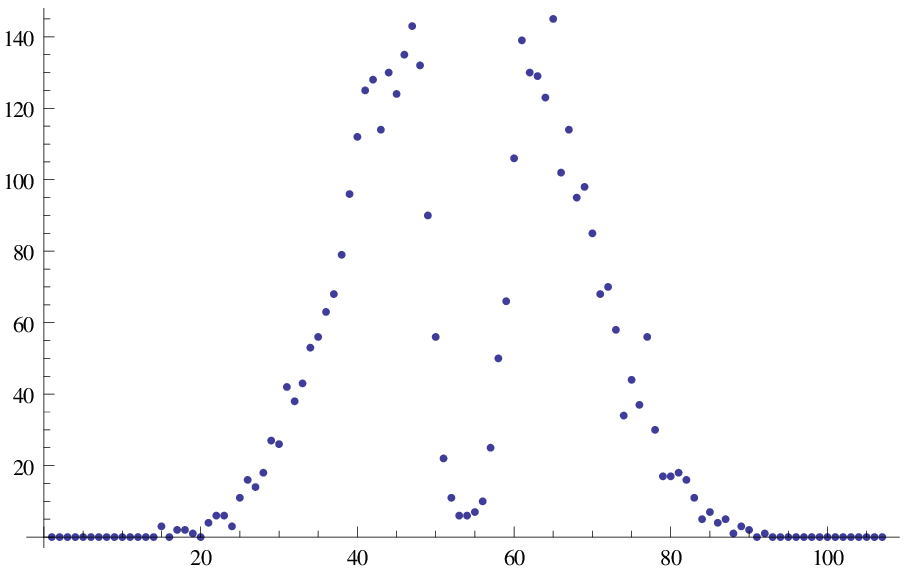}}{IV}\\
 \bigskip
\stackunder[5pt]{\includegraphics[width = 1.3in]{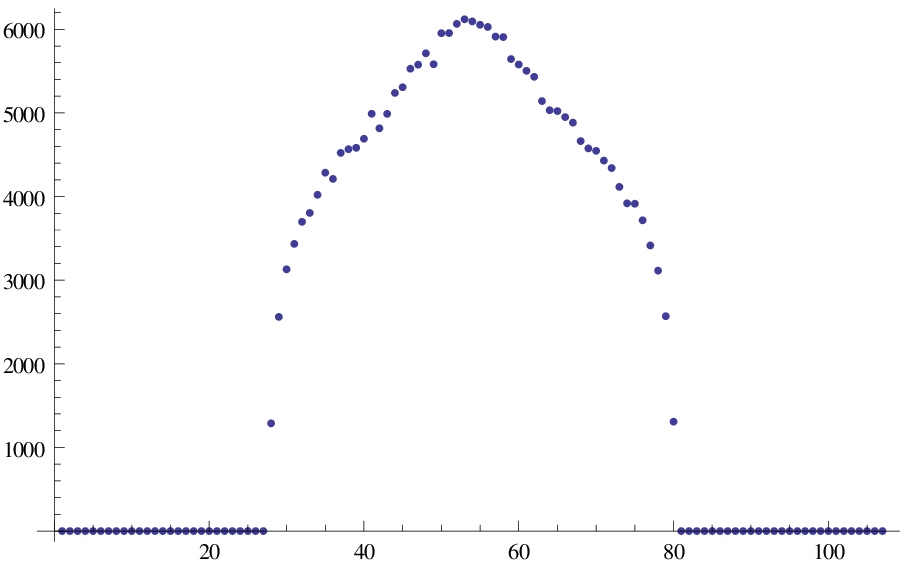}}{I} &
\stackunder[5pt]{\includegraphics[width = 1.3in]{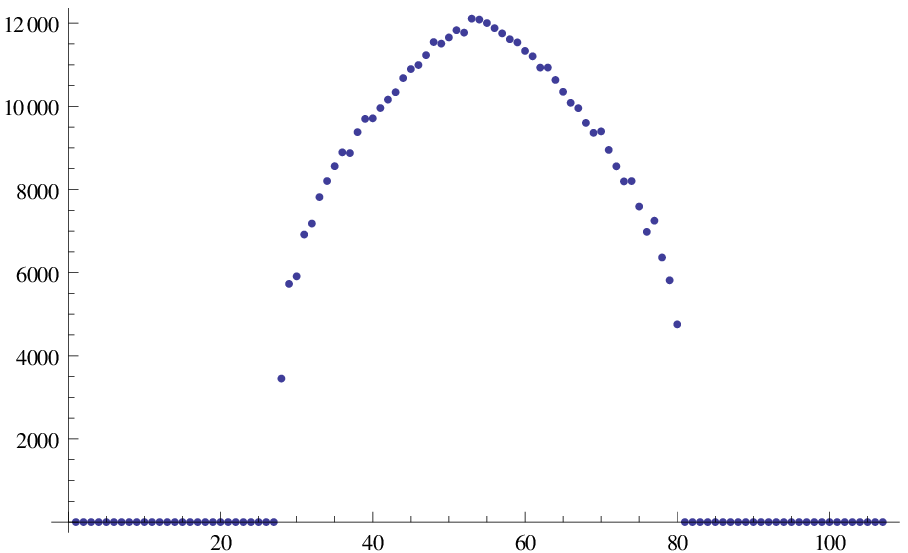}}{II'} &
\stackunder[5pt]{\includegraphics[width = 1.3in]{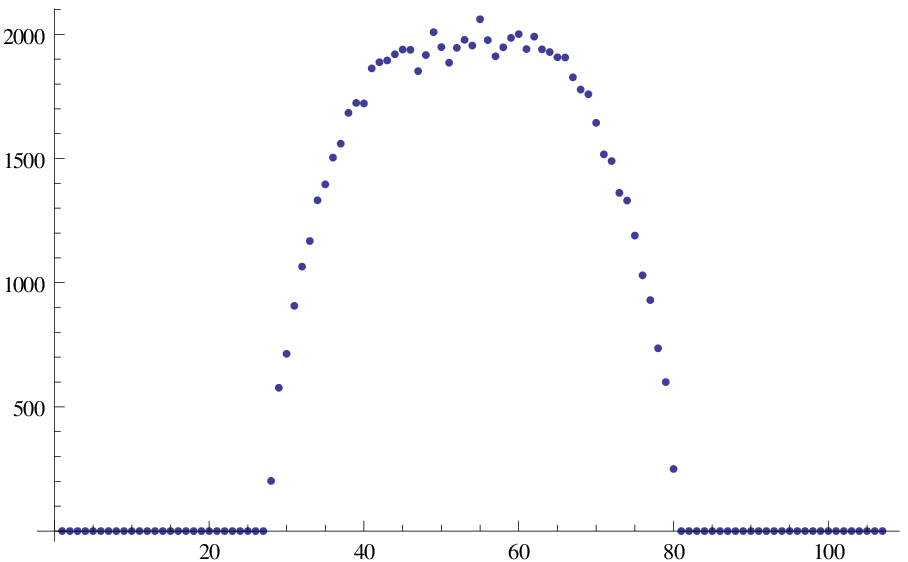}}{III, annihilations} &
\stackunder[5pt]{\includegraphics[width = 1.3in]{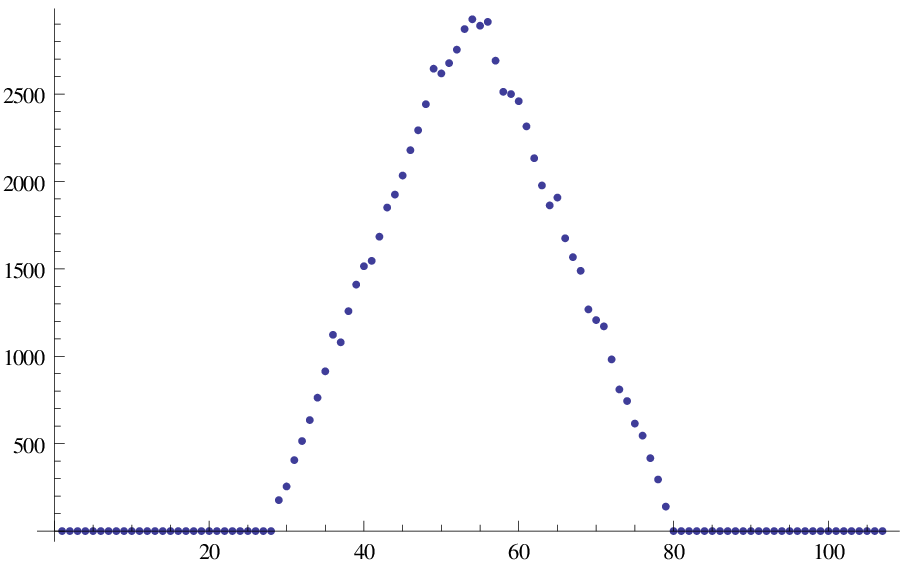}}{IV}\\
 \bigskip
\stackunder[5pt]{\includegraphics[width = 1.3in]{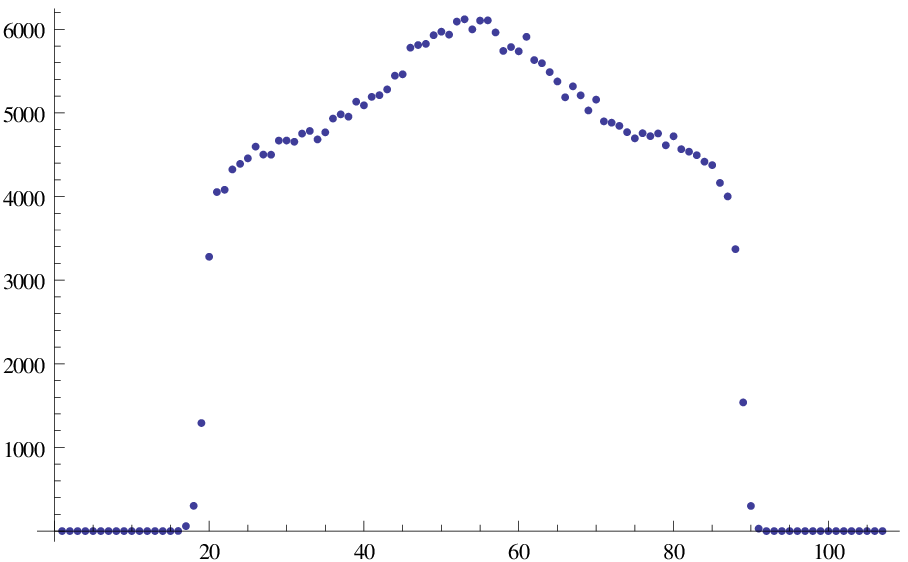}}{I} &
\stackunder[5pt]{\includegraphics[width = 1.3in]{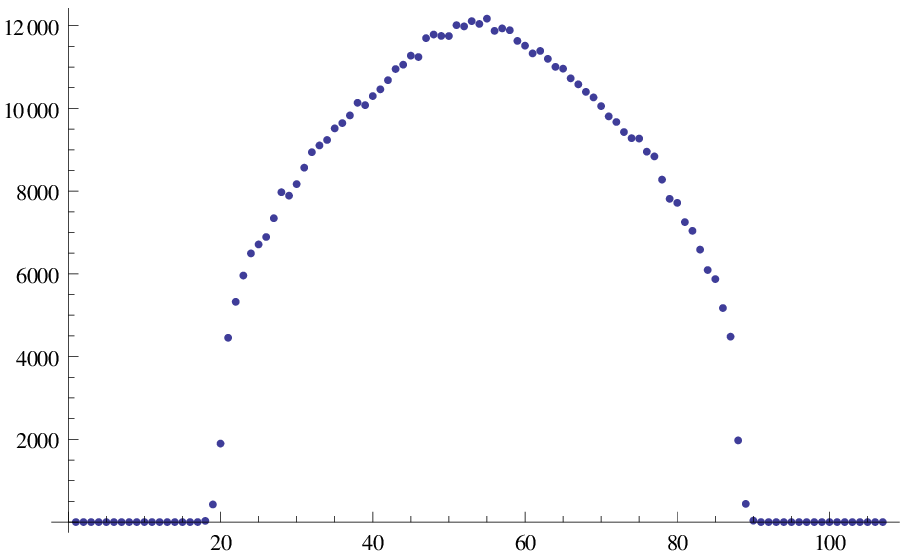}}{II'} &
\stackunder[5pt]{\includegraphics[width = 1.3in]{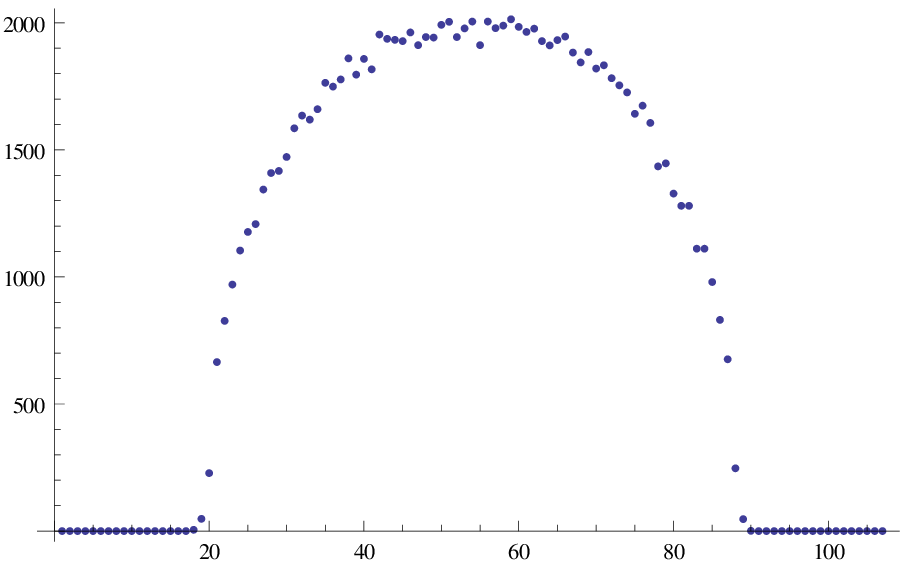}}{III, annihilations} &
\stackunder[5pt]{\includegraphics[width = 1.3in]{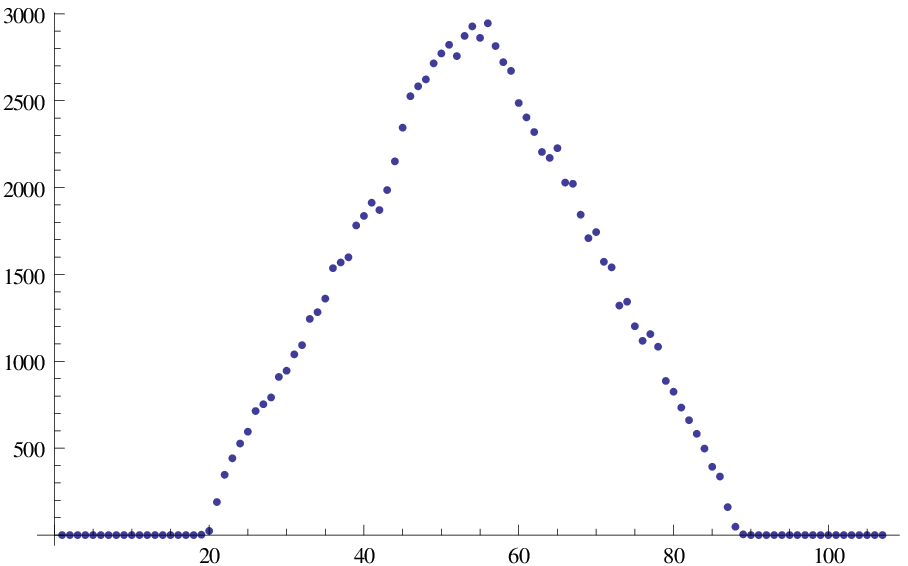}}{IV}\\
\end{tabular}

\caption{Density sections of actions at $(b,q)=(0,1)$: top two rows: horizontal and diagonal cuts for $p=\frac{1}{8}$, bottom two rows: horizontal and diagonal cuts for $p=\frac{7}{8}.$}

\end{figure}

\section{Conclusion}

\noindent One can view the results here as a study of the entropy geometry of the bounded 19-vertex model. All the randomness entering into the dynamic model does so via the four elementary actions. The spatial distributions of these actions reveal where the configurations are loose and where rigid and thereby where the entropy in the ensemble of the configurations is localized. An abrupt change in these distributions indicates a spatial phase transition of some kind. The actions are the smallest allowed perturbations of the configurations yet they seem to lead generically (at least in all our three parameters) to a striking macroscopic phenomenon, the emergence of equilibrium limit shapes. This is not a carry over from the Ice model as we have seen; Ice dynamics can be suppressed yet the phase segregation prevails. Whether any of the other (sub)actions are actually irrelevant for the limit shape formation and only influence the interior statistics of the temperate phase remains an open problem and so do the exact shapes.

\end{document}